\documentclass{article}

\usepackage{arxiv}

\usepackage[utf8]{inputenc} % allow utf-8 input
\usepackage[T1]{fontenc}    % use 8-bit T1 fonts
\usepackage{hyperref}       % hyperlinks
\usepackage{url}            % simple URL typesetting
\usepackage{booktabs}       % professional-quality tables
\usepackage{amsfonts}       % blackboard math symbols
\usepackage{nicefrac}       % compact symbols for 1/2, etc.
\usepackage{microtype}      % microtypography
\usepackage{lipsum}

\usepackage{amsbsy}
\usepackage{amssymb}
\usepackage{amsthm}
\usepackage{mathrsfs}
\usepackage{mathtools}
\usepackage{color}
\usepackage{graphicx} % Required for including images

\usepackage{t1enc}
\usepackage{amsmath}
\usepackage{latexsym}
\usepackage{natbib}
\usepackage{epsfig}
\usepackage{textcomp}
\usepackage{alltt}
\usepackage{graphicx}
\usepackage{rotating}
\usepackage{dcolumn}
\usepackage{bigints}
\newcommand{\nothere}[1]{}

\title{Estimation of separable direct and indirect effects in continuous time}

\author{
  Torben Martinussen\\
Department of Biostatistics,
University of Copenhagen \\
tma@sund.ku.dk\\
\AND
Mats Julius Stensrud\\
Department of Mathematics,
Ecole Polytechnique Fédérale de Lausanne\\
m.j.stensrud@gmail.com
}

\begin{document}
\maketitle

\begin{abstract}
Many research questions involve time-to-event outcomes that can be prevented from occurring due to competing events. In these settings, we must be careful about the causal interpretation of classical statistical estimands. In particular, estimands on the hazard scale, such as ratios of cause specific or subdistribution hazards, are fundamentally hard to be interpret causally. Estimands on the risk scale, such as contrasts of cumulative incidence functions, do have a causal interpretation, but they only capture the total effect of the treatment on the event of interest; that is, effects both through and outside of the competing event. To disentangle causal treatment effects on the event of interest and competing events, the separable direct and indirect effects were recently introduced. Here we provide new results on the estimation of direct and indirect separable effects in continuous time. In particular, we derive the nonparametric influence function in continuous time and use it to construct an estimator that has certain robustness properties. 
We also propose a simple estimator based on semiparametric models for the two cause specific hazard functions.
%, and define multiply robust estimators that can be implemented with standard statistical software. 
%We also consider alternative weighted estimators that are simple to implement. 
We describe the asymptotic properties of these estimators, and present results from simulation studies, suggesting that the estimators behave satisfactorily in finite samples. Finally, we re-analyze the prostate cancer trial from Stensrud et al (2020). 
%We provide sufficient conditions for identification of the separable effects using local independence graphs.
\end{abstract}
%Yet, the theory on separable effects has hitherto focused on interpretation and identification, and estimators have only been described in a discrete-time setting.

% keywords can be removed

\section{Introduction}
In survival analysis, the event of interest can be prevented from occurring due to a competing event. The presence of competing events requires us to be careful about the interpretation of classical statistical estimands \citep{robins1986new, young2018choice}. In particular, it is well-established that estimands on the hazard scale, such as cause specific hazard or subdistribution hazard ratios, do not have a causal interpretation unless we impose strong assumptions that usually are unreasonable \citep{robins1986new, hernan2010hazards,martinussen2018subtleties, young2018choice}. Yet the cumulative incidence function, which is defined on the risk scale \citep{andersen2012competing}, has a causal interpretation as the total effect on the event of interest \citep{young2018choice}. 

However, the total effect does not inform us about the mechanisms by which the treatment exerts effects on the event of interest. To illustrate this, suppose that we perfectly executed a randomized experiment in which 1000 patients received a cancer drug and 1000 patients received control. After 5 years, 250 patients died in the treatment arm and 500 died in the control arm, and therefore the drug was successfully shown to reduce mortality after 5 years. However, the drug was excreted in the kidneys, and to assess a potential side effect, the investigators did an additional analysis in which kidney failure was the primary outcome. They found 250 kidney failures in the treatment arm and 100 kidney failures in the placebo arm after 5 years. Two scientists debated the treatment effect on kidney events. As the drug was known to be excreted in the kidneys, the first scientist suspected that the increase in kidney events was a biological side effect. The second scientist doubted this explanation, and claimed that the increase in kidney events occurred because the drug reduced mortality, and hence more subjects were at risk of developing kidney events in the treatment arm. Thus, even though the scientists could identify the cumulative incidences of mortality and kidney events, they were unable to agree on the causal mechanism by which treatment causes kidney events. 

It has sometimes been suggested that the marginal distribution function, also called the net risk, can be used to assess the causal effect of treatment on the event of interest outside of its effect on the competing event. However, interpreting this estimand requires us to consider a hypothetical intervention to prevent the competing event from occurring, that is, to consider a controlled direct effect \citep{robins1992identifiability, young2018choice}. This is problematic because the hypothetical intervention to prevent the competing event, which is death in our conceptual example above, is usually infeasible in practice, and therefore the causal estimand is ill-defined and is not scientifically interesting. In particular, the controlled direct effect cannot solve the problem raised by the two scientists in the example above. 

To describe the mechanism by which the treatment exerts effects on the outcome of interest, \citet{stensrud2019separable} recently introduced the separable direct and indirect effects, motivated by a treatment decomposition idea introduced by \citet{robins2010alternative} in a mediation setting (see also \citet{didelez2018defining}). The separable direct effect is the treatment effect on the event of interest outside of its effect on the competing event, and the separable indirect effect is the effect on the outcome of interest only through the competing event \citep{stensrud2019separable}. These effects add up to the total effect, that is, the conventional cumulative incidence function, and thereby the separable effects describe the mechanism by which the total effect arise. In particular, the discussion between the scientists above could be resolved by considering the separable effects \citep{stensrud2019separable}.

So far the theory on the separable effects has been restricted to settings in discrete time \citep{stensrud2019separable} and focused on identification and interpretation. Here we define separable effects in continuous time and consider various estimators. Importantly, we derive the nonparametric influence function for the separable effects that leads to a robust estimator. We describe the asymptotic properties of the estimator, assess its properties in a simulation study, and re-analyze the example on prostate cancer from \citet{stensrud2019separable}.

The manuscript is organized as follows. In Section \ref{sec: obs data}, we describe the observed data structure, establish notation that will be used in subsequent sections and define the separable effects in continuous time. In Section \ref{sec: derive influence function}, we derive the nonparametric influence function of the separable direct effect in continuous time, propose an estimator based on the influence function, often referred to as the one-step estimator, and describe its robustness to model mis-specification. In Section \ref{sec: large sample properties}, we study the asymptotic properties of the estimator when Cox proportional hazard models are used for estimation.
We also describe the asymptotic properties of the one-step estimator. In Section 
\ref{sec: influence function indirect effect}, we give the efficient influence function of the separable indirect effect.
In Section \ref{sec: simulations}, we assess the performance of the estimators in a simulation study and re-analyse the prostate cancer data from \cite{stensrud2019separable}. In section \ref{sec: conclusion}, we provide a discussion. Detailed calculations are given in the Appendix.

\section{Data structure and notation}
\label{sec: obs data}
Suppose that we observe data from a randomized experiment in which $i=1,\ldots ,n$ individuals are assigned a dichotomous treatment $A_i$. For each individual $i$, we measure data on a vector of covariates, $W_i$, before treatment assignment, and each $i$ has an event time $T_i$ of type $\epsilon_i \in \{1,2\}$, where $\epsilon_i=1$ denotes the event of interest ($Y$) and $\epsilon_i=2$ the competing event $D$. We will hereby suppress the individual $i$ subscript, because the random vector for each individual is assumed to be drawn independently from a distribution common to all subjects. Because of loss to follow-up, we only observe $\Delta=I(T\leq C)$ and $\tilde T=\mbox{min}(T,C)$, where $C$ denotes the right censoring time. 
We assume that $T$ and $C$ are conditionally independent given $(A,W)$.
To focus on the main ideas of this work and simplify the derivations, we initially develop the theory without censoring. Thus, we first assume that data for a given individual consist of the vector $X=(T,\epsilon,A,W)$. However, the results immediately extend to settings with censoring, where the censoring may depend on $A$ and $W$, as we describe in more detail in Section \ref{sec: derive influence function}. 

\subsection{Treatment decompositions}
\label{sec: decomposition}
To define our estimands of interest -- the separable effects -- we consider a decomposition of $A$ into two dichotomous components, $A_{Y}$ and $A_{D}$, analogous to \citet{stensrud2019separable}: the $A_Y$ component exerts all its effects on the event of interest ($Y$) outside of the competing event ($D$), and $A_D$ exerts all its effect on $Y$ through $D$. In the observed data, the treatment components are deterministically related in each individual, $A = A_{Y} = A_{D}$, but we will conceive a hypothetical experiment in which these components are assigned different values. This decomposition assumption motivates the definition of our separable direct and indirect effects. However, the separable effects can still be meaningful, even if a physical decomposition of the treatment is possible, e.g.\ if we can conceive a hypothetical treatment that operate in the same way as the $A_Y$ component of $A$, but does not exert effects on $A_D$ \citep{stensrud2019separable, stensrud2019generalized}.

\subsection{Definition of separable effects}
We use superscripts to denote counterfactuals, such that $T^{a}$, is the event time when, possibly contrary to fact, $A$ is set to $a$, and $\epsilon^a$ indicates whether an event of interest $Y$ ($\epsilon^a=1$) or a competing event $D$ ($\epsilon^a=2$) occurred at $T^{a}$. Similarly,  $T^{a_Y,a_D}$ and $\epsilon^{a_Y,a_D}$ denote counterfactual values under an intervention that sets $A_Y$ to $a_Y$ and $A_D$ to $a_D$.

Let 
\begin{align*}
P_1(t, a_Y,a_D)\equiv P(T^{a_Y,a_D}\leq t,\epsilon^{a_Y,a_D}=1).
\end{align*}

Analogous to the results in \cite{stensrud2019separable} in discrete time, we can now define the separable direct effect of treatment $A$ at time $t$ as
\begin{align*}
   & P_1(t, 1,a_D) \text{ vs. } P_1(t, 0,a_D) \quad \text{for } a_D \in \{0,1\},
\end{align*}
and the separable indirect effect is defined as 
\begin{align*}
   & P_1(t, a_Y,1) \text{ vs. } P_1(t, a_Y,0) \quad \text{for } a_Y \in \{0,1\}.
\end{align*}

Note that pairs of separable direct and indirect effects sum to the total effect, which is equal to the classical cumulative incidence function, that is,
\begin{align*}
 \bigl\{ P_1(t, 1,a)  - P_1(t, 0,a) \bigr\}
+   \bigl\{P_1(t, 1-a,1) - P_1(t, 1-a,0)  \bigr\}
=  P_1(t, 1,1) - P_1(t, 0,0)
\end{align*}
for $a\in \{0,1\}.$

%We are interested in comparing 
%$$
%P_1(t, a^{*},a)\equiv P(T^{a^{*},a}\leq t,\epsilon^{a^{*},a}=1)
%$$
%for different values of $a^{*}$ and $a$, specifically we wish to compute an estimator of 
\subsection{Identifiability conditions}
\label{sec: identification}
Consider the additive separable direct effect,
$$
\delta_1(t,a_D)=P_1(t, 1,a_D) - P_1(t, 0,a_D) \text{ for } a_D \in \{0,1\}.
$$
To identify $\delta_1(t,a_D)$ from the observed data, where $A$ and the components $A_Y$ and $A_D$ are deterministically related, we impose the following conditions, which are continuous time analogues to the conditions in \citet{stensrud2019separable}, see also \citet{robins2010alternative}.

We assume conditional exchangeability, that is, 
\begin{align*}
& (T^a, \epsilon^a) \perp\!\!\!\perp A \mid W \text{ for } a \in \{0,1\},
\end{align*}%
which is a classical exchangeability condition that is expected to hold when treatment $A$ is randomly assigned. 

Second, we assume consistency, such that if an individual has observed treatment $A=a$, then 
\begin{align*}
&(T^a, \epsilon^a) = (T, \epsilon),
\end{align*}
for $a\in\{0,1\}$. The consistency assumption ensures that the observed outcome is equal to the counterfactual outcome for any individual that has observed data history consistent with a counterfactual scenario.

Third, positivity such that % for all $l \in \mathcal{L}$,
\begin{align}
 & f(W=w)>0\implies   \notag \\
 & \quad \Pr (A=a\mid  W=w)>0 \text{ for } a\in\{0,1\}, %  \text{ and }  l \in \mathcal{L}, 
\label{eq: positivity of A} \\ 
  & f(T > t, W=w) > 0  \implies \nonumber\\ 
 & \quad  \Pr(\Tilde{T} > t, A=a| W=w)>0 \text{ for } a\in\{0,1\} \text{ and for all } t <t^*, %\label{eq: positivity of A part 2}
%& \quad \Pr (A=a\mid  L)>0\text{ w.p.1} \text{ for } a\in\{0,1\},
\label{eq: positivity of A part 2}
\end{align}%
where $t^*$ denotes the end of follow-up and $f$ generically denotes a density function.
Note that \eqref{eq: positivity of A} is the usual positivity condition under interventions on $A$ and \eqref{eq: positivity of A part 2} ensures that among those event-free through each follow-up time, there exist individuals with $A=1$ and individuals with $A=0$ that are uncensored. 

Finally, we impose dismissible component conditions. To introduce these conditions, let $\lambda_j(t|A=a,W=w)$ denote the conditional cause specific hazard function with $j=1, 2,$ denoting the $j$th cause, and let $\Lambda_{j}(s|A=a,W=w)=\int_0^t\lambda_j(s|A=a,W=w)\, ds$ be the corresponding cumulative conditional cause specific hazard function. Similarly, let $\lambda_j^{a}(t|W=w)$ and $\lambda_j^{a_Y,a_D}(t|W=w)$ be the counterfactual conditional cause specific hazard functions under interventions on $A$ and joint interventions on $A_Y$ and $A_D$, respectively. Then, the dismissible components conditions are

\begin{align*}
\mathbf{\Delta1:\; }& \lambda_1^{a_Y,a_D=1}(t|W=w) = \lambda_1^{a_Y,a_D=0}(t|W=w), \quad a_Y \in \{0,1\},
\end{align*}%
at all $t$, which states that a counterfactual hazards of the event of interest ($j=1$) are equal under all values of $A_{D}$, and
\begin{align*}
\mathbf{\Delta2:\; }& \lambda_2^{a_Y=1,a_D}(t|W=w) = \lambda_2^{a_Y=0,a_D}(t|W=w), \quad a_D \in \{0,1\},
\end{align*}%
at all $t$, which states that a counterfactual hazard functions of the competing event ($j=2$) are equal under all values of $A_{Y}$. %The dismissible component conditions are analogous to identification conditions from Shpitser \cite{shpitser2013counterfactual} on path specific effects.

\subsection{Functionals of counterfactual and observed data}
%Let $\lambda_T^j(t|A=a,W=w)$ denote the conditional cause specific hazard function with $j=1, 2,$ denoting the $j$th cause, and let $\Lambda^{j}_T(s|A=a,w)=\int_0^t\lambda_T^j(s|A=a,W=w)\, ds$ be the corresponding cumulative conditional cause specific hazard function. Furthermore, 
Let  
\begin{align*}
P_1(t,a_Y,a_D,w)&=\int_0^te^{-\Lambda_{1}(s|A=a_Y,w)-\Lambda_{2}(s|A=a_D,w)}d\Lambda_{1}(s|A=a_Y,w),\\
\delta_1(t,a_D,w)&=   P_1(t,1,a_D,w)-P_1(t,0,a_D,w). 
\end{align*}

Note that under the treatment decomposition assumption \citep{robins2010alternative, stensrud2019separable, stensrud2019generalized} and  the identification conditions in Section \ref{sec: identification}, the cumulative incidence function for $Y$ under treatment $a$ conditional on $W=w$ is given by 
$P_1(t,a,a,w)$,  which is also denoted by $F_1(t|a,w)$.
Then a continuous-time equivalent to the G-formula \citep{robins1986new} in \citet{stensrud2019separable} is
\begin{align*}
P(T^{a_Y,a_D}\leq t,\epsilon^{a_Y,a_D}=1)
=\int P_1(t,a_Y,a_D,w) f(W=w)\, dw,
\end{align*}
which allows us to identify our parameter of interest, $\delta_1(t,a_D)=E\{\delta_1(t,a_D,W)\}$, from the observed data.

\subsection{Estimation using classical regression models}
Suppose we were willing to postulate (semi)parametric models for the cause specific hazard functions, such as Cox proportional hazards models. Then it would be straight forward to estimate
 $\delta_1(t,a)$ using
$$
\hat\delta_1(t,a)=\hat P_1(t,1,a)-\hat P_1(t,0,a),
$$
where
\begin{align*}
\hat P_1(t,1,a)=&n^{-1}\sum_i\left \{\int_0^te^{-\hat\Lambda_{1}(s|A=1,W_i)-\hat\Lambda_{2}(s|A=a,W_i)}d\hat\Lambda_{1}(s|A=1,W_i)\right\}, \\
 \hat P_1(t,0,a)= &n^{-1}\sum_i\left \{\int_0^te^{-\hat\Lambda_{1}(s|A=0,W_i)-\hat\Lambda_{2}(s|A=a,W_i)}d\hat\Lambda_{1}(s|A=0,W_i)\right\},
\end{align*}
because the terms in $P_1(t,1,a)$ and $P_1(t,0,a)$ can easily be estimated using Cox-models for the two cause specific hazard functions. That is, if $\lambda_j(t|a,w)=\lambda_{j0}(t)e^{\beta_j^T l}$, with $l=(a,w^T)^T$, then $\hat \Lambda_{j}(s|l)$ is  obtained from $\hat \Lambda_{0j}(t)e^{\hat \beta_j^T l}$, which can be estimated from a Cox regression analysis. Asymptotic properties can also be derived using the approach of \cite{10.2307/25734118}, which we return to in Section  \ref{Asymp_Cox}. 

However, using such semiparametric regression models for the cause specific hazard functions may lead to biased results if these models are misspecified. Therefore we will provide more general results, based on semiparametric theory \citep{Vaart:1998aa, Vdl-Robins_book}, which leads us to estimators with desirable properties such as semiparametric efficiency and certain robustness  \citep{bang2005doubly}. The primary tool to finding these estimators is to derive the so-called efficient influence function, see \cite{Vaart:1998aa}, which we do in Section \ref{sec: derive influence function}.

\section{The  efficient influence function and estimation of the separable direct effect}
\label{sec: derive influence function}
%\label{sec: estimation}
In this section we give the efficient influence function for the target parameter parameter
$$
\psi_t(P)=\delta_1(t,1)=E\{P_1(t,1,1,W)\}-E\{P_1(t,0,1,W)\},
$$
where we use $P$ to denote the probability measure from which we observe $Z=(T,\epsilon,A,W)$. Note that it is possible to recode the treatment variable $A$, that is,  interchanging the two levels 0 and 1, and thus we can restrict our attention to $\delta_1(t,1)$ without loss of generality. The corresponding separable indirect effect is $P_1(t,0,1)-P_1(t,0,0)$, which we decribe in more detail in Section \ref{sec: influence function indirect effect}.

We impose no structure on $P$ and show in 
the Appendix that the efficient influence function is
\begin{align*}
%\label{Del1_infl}
&\tilde \psi(t,Z)\notag \\
=&\bigl\{ N_1(t)-P_1(t,1,1,W)\bigr \}\frac{I(A=1)}{P(A=1|W)}    
-\bigl \{ \int_0^t\frac{e^{-\Lambda_2(s|1,W)}}{e^{-\Lambda_2(s|0,W)}}dN_1(s)-P_1(t,0,1,W)\bigr \} \frac{I(A=0)}{P(A=0|W)}
\notag\\
&-\int_0^t   \bigl \{P_1(t,0,1,W)-P_1(u,0,1,W)\bigr \} \biggl [\frac{dM^T_{2}(u|0,W)}{P(T>u|0,W)}\frac{I(A=0)}{P(A=0|W)} 
- \frac{dM^T_2(u|1,W)}{P(T>u|1,W)}\frac{I(A=1)}{P(A=1|W)}\biggr ] \notag \\
&+\delta_1(t,1,W)- E\bigl\{ \delta_1(t,1,W)\bigr \},
\end{align*}
where $N_j(t)=I(T\leq t,\epsilon=j)$ is the $j$th specific counting process and $M^T_j(t|a,w)$ is the corresponding counting process martingale given $A=a, W=w$, i.e.,  $M^T_j(t|a,w)=N_j(t)-\int_0^tI(s\leq T)d\Lambda_j(s|a,w)$.
 We may further rewrite the efficient influence function in terms of the counting process martingales, see the Appendix for further details,
\begin{align}
\label{Del1_infl_expr3}
 \tilde \psi(t,Z)= 
  &\int h_1(s,t,A,W)dM^T_1(s|A,W)
+\int h_2(s,t,A,W)dM^T_2(s|A,W) \nonumber\\
&+\delta_1(t,1,W)- E\bigl\{ \delta_1(t,1,W)\bigr \}
\end{align}
where
\begin{align*}
h_1(s,t,A,W)=&I(s\leq t)g(A,W)\frac{e^{-\Lambda_2(s|1,W)}}{e^{-\Lambda_2(s|A,W)}}\biggl\{ 1-  \frac{\{F_1(t|A,W)-F_1(s,|A,W)\bigr \}}{P(T>s|A,W)} \biggr\},\\
h_2(s,t,A,W)=&I(s\leq t)\frac{g(A,W)}{P(T>s|A,W)} \biggl \{P_1(t,0,1,W)-P_1(s,0,1,W)-\\
&\frac{e^{-\Lambda_2(s|1,W)}}{e^{-\Lambda_2(s|A,W)}}\bigl \{F_1(t|A,W)-F_1(s,|A,W)\bigr \} \biggr \},\\
g(A,W)=&\frac{A}{P(A=1|W)}-\frac{1-A}{P(A=0|W)}.
\end{align*}

We remind the reader that $C$ denotes the potential censoring time, $\tilde T=T\wedge C$  and  $\Delta=I(T\leq C)$. We can immediately generalize \eqref{Del1_infl_expr3} to allow for censoring, as we have imposed no structure on $P$ \citep[formula 10.76]{tsiatis-book2006}: the efficient influence function based on the observed data $D=(\tilde T,\Delta,\Delta\epsilon,A,W)$ is given by
\begin{equation}
    \label{Tsiatis-fomula}
\psi(t,D)=\frac{ \tilde \psi(t,Z)\Delta}{K_C(T|A,W)}+\int_0^{\infty}\frac{L(s,A,W)}{K_C(s|A,W)}dM_C(s|A,W),
\end{equation}
 where we let $\Lambda_C(s|A,W)=\int_0^s\lambda_C(u|A,W)\, du$  denote the cumulative censoring hazard function, $K_C(s|A,W)=e^{-\Lambda_C(s|A,W)}$ is the corresponding  survival function, and
 $$
 M_C(t|A,W)=N_C(t)-\int_0^t I(s\leq \tilde T)d\Lambda_C(s|A,W)
 $$
 is the martingale associated with the censoring counting process $N_C(t)=I(\tilde T\leq t,\Delta=0)$ using the filtration where we include $A$ and $W$.
 In \eqref{Tsiatis-fomula},
 \begin{align}
 \label{L-fun}
 L(s,A,W)=E\bigl\{ \tilde \psi(t,Z) |T>s, A,W \bigr\}.
 \end{align}
 Following Lemma A.2 of \citet{Lu_Tsiatis_08}, this is easily generalized to the competing risk setting considered here, we can express \eqref{Tsiatis-fomula} in terms of the counting process martingales based   on the observed data.
 We get that the efficient influence function based on the observed data can be written as 
 \begin{align}
\label{Obs_infl_expr3}
\psi(t,D)= \phi(t,D)
+\delta_1(t,1,W)- E\bigl\{ \delta_1(t,1,W)\bigr \},
\end{align}
where 
$$
 \phi(t,D)= \int \frac{h_1(s,t,A,W)}{K_C(s|A,W)}dM_1(s|A,W)
+\int \frac{h_2(s,t,A,W)}{K_C(s|A,W)}dM_2(s|A,W)
$$
and where
 $M_j(t|a,w)$, $j=1,2$, are the observed counting process martingales given $A=a, W=w$, i.e.,  for $j=1,2$,
 $$M_j(t|a,w)=\Delta I(\tilde T\leq t,\epsilon=j)-\int_0^tI(s\leq \tilde T)d\Lambda_j(s|a,w).
 $$

%\section{Estimation}

The  one-step estimator based on the efficient influence function is thus given by
\begin{equation}
\label{EIF-est}
     \hat \delta_{1e}(t,1)=n^{-1}\sum_{i=1}^n \biggl\{\hat \delta_1(t,1,W_i)+ \hat { \phi}(t,D_i)\biggr\},
\end{equation}
where  $\hat \delta_1(t,1,W_i)$ is defined analogously to $ \delta_1(t,1,W_i)$, except that the unknown quantities are replaced with estimated  counterparts, and similarly for $ \hat { \phi}$.
This part requires working models, and in Section \ref{sec: robustness} we describe how robust the resulting estimator is to mis-specification of these working models. Note also that the one-step estimator is equal to the simple estimator $ \hat \delta_{1}(t,1)$ plus an augmentation term.
   
\subsection{Robustness}
\label{sec: robustness}
  We argue now that the estimator $\hat \delta_{1e}(t,1)$ given in
\eqref{EIF-est} has certain robustness properties unlike  the initial estimator $\hat \delta_1(t,1)$ based on Cox-models for the cause specific hazard functions. Consider first the setting where there is no censoring, i.e., we are then basing estimation on the efficient influence function $\tilde \psi(t,Z)$ given in \eqref{Del1_infl_expr3}. Let  $H$ denote the unknown parameters that goes into the efficient influence function $\tilde \psi(t,Z)$, i.e. $H=\{\Lambda_1(\cdot|A,W),\Lambda_2(\cdot|A,W),P(A=1|W)\}$.
Our estimator in this case is then the solution to $0=n^{-1}\sum_i \tilde \psi(t,Z_i,H_n)$, where $H_n$ is an estimator of $H$. We show in the Appendix that the resulting estimator is consistent if two out of the three possible working models are correctly specified. 
%If the propensity score model is correctly specified then we obtain consistency if one of %the cause specific hazard models is correctly specified but not necessarily both. 
Now consider the setting where we allow for censoring.
Let $G$ denote the unknown parameters that goes into the efficient influence function, i.e. $G=\{\Lambda_1(\cdot|A,W),\Lambda_2(\cdot|A,W),P(A=1|W),\Lambda_C(\cdot|A,W)\}$ so there are now four models, which we denote (i) to (iv) in the order indicated in the definition of $G$.
We show  in the Appendix that $\hat\delta_{1e}(t,1)$ is consistent 
if the following models are correctly specified: (i) and (ii), or (i), (iii) and (iv), or (ii), (iii) and (iv). Hence, in a randomized study with the censoring being independent of $W$ then we obtain consistency if one of the two cause specific hazard models is correctly specified, but not necessarily both.

\section{Large sample properties}
\label{sec: large sample properties}

\subsection{Properties under proportional cause specific hazards}
\label{Asymp_Cox}
In this subsection we assume that the cause specific hazard functions are on Cox proportional hazards form, i.e. that these models are correctly specified.
\nothere{Let
$$
P_1(t,a_Y,a_D,w)=\int_0^te^{-\Lambda_{1}(s|A=a^{*},w)-\Lambda_{2}(s|A=a,w)}d\Lambda_{1}(s|A=a^{*},w)
$$
and let $\hat  P_1(t,a_Y,a_D,l)$ be as $P_1(t,a_Y,a_D,l)$ with estimates inserted. To this end, we model
the cause specific hazard functions using the Cox-model.
}
Specifically,
let 
$$
\Lambda_j(t|a,w)=\Lambda_{0j}(t)e^{ \beta_A^j a+\beta_W^j w},\; j=1,2,
$$
and
$$
\hat P_1(t,a_Y,a_D)=n^{-1}\sum_i\hat P_1(t,a_Y,a_D,W_i),
$$
where 
$\hat P_1(t,a_Y,a_D,W_i)$ is calculated using the estimates from fitting separate Cox regression models with the event of interest and the competing event as dependent variable.
We show in the Appendix that
$$
n^{1/2}\left\{\hat P_1(t,a_Y,a_D)- P_1(t,a_Y,a_D)\right \}=\sum_i \epsilon_i^{P_1}(t,a_Y,a_D),
$$
where  	$\epsilon_i^{P_1}(t,a_Y,a_D)$ are zero-mean iid terms 
(i.e.\  the influence function) so that $\hat P_1(t,a_Y,a_D)$ is a
RAL estimator \citep{tsiatis-book2006} as long as the specified Cox proportional hazards models for the cause specific hazard functions are correctly specified.
In the Appendix, we also give further details on how to
estimate the influence function enhancing estimation of the variance of the estimator. 

\subsection{Non-parametric properties}
\nothere{Let $G$ denote the unknown parameters that goes into the efficient influence function, i.e. $G=\{\Lambda_1(\cdot|A,W),\Lambda_2(\cdot|A,W),P(A=1|W),\Lambda_C(\cdot|A,W)\}$.

Further}
Let
$$\tilde \phi(t,D,G)=\delta_1(t,1,W,G)+\phi(t,D,G).
$$
so that efficient influence function $\psi(t,D,G)$ is re-expressed as $\tilde \phi(t,D,G)-\delta_{1}(t,1)$.
If $G$ is known then 
\begin{align*}
    0&=n^{-1}\sum_i \bigl\{\tilde\phi(t,D_i,G)-\hat\delta_{1e}(t,1)\bigr\}\\
    &=n^{-1}\sum_i\bigl\{\tilde \phi(t,D_i,G)-\delta_{1}(t,1)\bigr\}-\{\hat\delta_{1e}(t,1)-\delta_{1}(t,1)\}
\end{align*}
from which we see that $\hat\delta_{1e}(t,1)$ has influence function
$\psi(t,D,G)$, ie, the efficient influence function. Thus, in this case,  $\hat\delta_{1e}(t,1)$ is a semiparametrically efficient RAL estimator \citep{tsiatis-book2006}.
In reality $G$ is not known and needs to be estimated. Let $G_n$ be such an estimator of $G$.
The proposed estimator $\hat\delta_{1e}(t,1)$ solves
$$
 0=n^{-1}\sum_i \bigl\{\tilde\phi(t,D_i,G_n)-\hat\delta_{1e}(t,1)\bigr\}
$$
and, therefore,
\begin{align*}
n^{1/2}\{\hat\delta_{1e}(t,1)-\delta_{1}(t,1)\}&=n^{-1/2}\sum_i\bigl\{\tilde \phi(t,D_i,G)-\delta_{1}(t,1)\bigr\}+n^{-1/2}\sum_i\bigl\{\tilde \phi(t,D_i,G_n)-\tilde \phi(t,D_i,G)\bigr\}\\
&=n^{-1/2}\sum_i\psi(t,D_i,G)+n^{-1/2}\sum_i\bigl\{\tilde \phi(t,D_i,G_n)-\tilde \phi(t,D_i,G)\bigr\}\\
&=n^{-1/2}\sum_i\psi(t,D_i,G)+En^{1/2}\bigl\{\tilde \phi(t,D,G_n)-\tilde \phi(t,D,G)\bigr\}+o_p(1),
\end{align*}
following \citet{10.2307/25734118}.
The expectation on the right hand side of the latter display is taken w.r.t. to $D$. Based on correctly specified models $G_n$ one may then derive the true analytical form of the influence function of $\hat\delta_{1e}(t)$ in which case $\hat\delta_{1e}(t,1)$ is still a semiparametrically RAL estimator.
The expression of the influence function depends on the specific chosen working models $G_n$ similar to the development in Section \ref{Asymp_Cox}, however. Instead we recommend using the non-parametric bootstrap procedure to estimate the variance of  $\hat\delta_{1e}(t)$ similar to what has been advocated in for instance \cite{van_der_Laan_Rubin_2006} and
\cite{van_der_Laan_2011}.
Later, we discuss an alternative approach for estimation of the variance of $\hat\delta_{1e}(t)$.

\nothere{Keep in mind that $E\{\tilde\phi(t,D,G_0)\}=\delta_{1}(t,1)$ involving only the nuisance parameters $F=\{\Lambda_1(\cdot|A,W),\Lambda_2(\cdot|A,W)\}$.

It therefore  follows that $n^{1/2}\{\hat{\delta}_{1e}(t,1)-
\delta(t,1)\}$ converges to a Gaussian limit with zero mean, and with a variance $\sigma_e^2(t)$ where $\sigma_e(t)$    is consistently estimated by
$$
\hat \sigma_e(t)=n^{-1}\biggl [\sum_{i=1}^n \bigl\{
\psi(t,D_i,G_n)+\psi^{*}(t,D_i,H_n)
\bigr\}^2\biggr ]^{1/2},
$$
where 
$$
\psi^{*}(t,D,H)=
$$
}
   \section{The  efficient influence function of the separable indirect effect}
\label{sec: influence function indirect effect}

Analogous to the results in Section \ref{sec: derive influence function}, we get the efficient influence function for the separable indirect effect,
 $P_1(t,0,1)-P_1(t,0,0),$
which can be written as a function of the observed data,
 \begin{align}
\label{ID_ie}
\psi^{I}(t,D)= \phi^{I}(t,D)
+P_1(t,0,1,W)-P_1(t,0,0,W)- \bigl\{P_1(t,0,1)-P_1(t,0,0)\bigr \},
\end{align}
where 
$$
 \phi^{I}(t,D)= \int \frac{h_1^{I}(s,t,A,W)}{K_C(s|A,W)}dM_1(s|A,W)
+\int \frac{h_2^{I}(s,t,A,W)}{K_C(s|A,W)}dM_2(s|A,W),
$$
 with
\begin{align*}
h_1^{I}(s,t,A,W)=I(s\leq t)&g(A,W)\left\{1-\frac{e^{-\Lambda_2(s|1,W)}}{e^{-\Lambda_2(s|A,W)}}\right\}\biggl\{ 1-  \frac{\{F_1(t|A,W)-F_1(s,|A,W)\bigr \}}{P(T>s|A,W)} \biggr\},\\
h_2^{I}(s,t,A,W)=I(s\leq t)&\frac{g(A,W)}{P(T>s|A,W)} \biggl [
P_1(t,0,1,W)-P_1(s,0,1,W)-\\
&\left\{1-\frac{e^{-\Lambda_2(s|1,W)}}{e^{-\Lambda_2(s|A,W)}}\right\}\bigl \{F_1(t|A,w)-F_1(s,|A,W)\bigr \} \biggr ],
\end{align*}
and
$M_j(t|a,w)$, $j=1,2$, are the observed counting process martingales given $A=a, W=w$.

%\section{Numerical results}
%\label{sec: simulations}
%In this section, w

\section{Simulations}
\label{sec: simulations}
%\subsection*{Simulation study}

\subsection{Performance of $\hat{\delta}_1 (t,1)$}
We first consider the performance of the estimator $\hat\delta_1(t,1)$ that is based on using Cox-proportional hazards models for the two cause specific hazard functions. Clearly, this estimator is only consistent if the proportional cause specific hazards models are correctly specified. To generate data we used
the cause specific hazard functions 
\begin{align*}
\lambda_1(t|A=a,W=w)&=\lambda_{10}(t)e^{\beta_A^1a+\beta_W^1w}\\
\lambda_2(t|A=a,W=w)&=\lambda_{20}(t)e^{\beta_A^2a+\beta_W^2w}
\end{align*}
with $\lambda_{10}(t)=0.05$,  $\beta_A^1=-\log{(2)}$, $\beta_W^1=0.5\log{(2)}$,  and with 
$\lambda_{20}(t)=0.1$, $\beta_A^2=-0$, $\beta_W^2=0.5\log{(2)}$. Treatment indicator $A$ was generated with $P(A=1)=0.5$, and the covariate $W$ was uniform on $(0,1)$.
Censoring was generated according to the minimum of  7 and an exponentially distribution with mean 12. We then applied the estimators 
$\hat P_1(t,1,1)$, $\hat P_1(t,0,1)$ and $\hat \delta_1(t,1)=\hat P_1(t,1,1)-\hat P_1(t,0,1)$
and their corresponding standard error estimators, all calculated at time points 2, 4 and 6. Results are summarized in Table 1. Each entry in the table is based on 1000 replicates.

\bigskip

\centerline{Table 1 about here}

\bigskip

\noindent
Both the estimator and its corresponding standard error estimator behave satisfactorily (Table 1). At the early time points the coverage is slightly less than nominal for $P_1(t,1,1)$ and   $P_1(t,0,1)$ when $n=400$ but improves for $n=800$.

\subsection{Performance of the estimator $\hat\delta_{1e}(t,1)$}
\noindent
We now assess the performance of the estimator $\hat\delta_{1e}(t,1)$ given in \eqref{EIF-est}, which is derived from the efficient influence function.
We investigate the robustness properties of this estimator.
 We also report results for the simple estimator $\hat\delta (t,1)$
 The exposure $A$ is binary and the covariate $W$ is uniform on $(0,1)$.
 To be able to compute $\hat\delta_{1e}(t,1)$ we need working models for the two cause specific hazard models, the propensity score and the censoring hazard function. We used Cox proportional hazard models  for the two cause specific hazard models with main effects of $A$ and $W$, a logistic regression model with main effects of  $W$ for the propensity score, and a Cox proportional hazards model for the censoring hazard function with effect of $A$ only.
 Let $L=I(W>1/2)$,
$\lambda_{10}(t)=0.05$, $\lambda_{20}(t)=0.1$, $\beta_{1A}=-\log{(5)}$,
$\beta_{2A}=0$,        $\beta_{1W}=\log{(2)}$
$\beta_{2W}=0.5\log{(2)}$. 
Censoring times were generated as $C=\mbox{min}(\tilde C,12)$ where $\tilde C$ was generated using the hazard function $\lambda_{\tilde C}(t|A,W)$ specified below. 
We used a sample size of $n=400$ and
 simulated data from the following  different scanarios.

\begin{itemize}
    \item[A1]$\quad$ All models are correctly specified.
\begin{align*}
&\lambda_1(t|A=a,W=w)=\lambda_{10}(t)e^{\beta_A^1a+\beta_W^1w},\; 
\lambda_2(t|A=a,W=w)=\lambda_{20}(t)e^{\beta_A^2a+\beta_W^2w}\\
&P(A=1|W)=\mbox{expit}(0+\log{(2)}(W-0.5)),\; \lambda_{\tilde C}(t|A,W)=12.
\end{align*}

    \item[A2]$\quad$ All models are correctly specified except the censoring model.
\begin{align*}
&\lambda_1(t|A=a,W=w)=\lambda_{10}(t)e^{\beta_A^1a+\beta_W^1w},\; 
\lambda_2(t|A=a,W=w)=\lambda_{20}(t)e^{\beta_A^2a+\beta_W^2w}\\
&P(A=1|W)=\mbox{expit}(0+\log{(2)}(W-0.5)),\; \lambda_{\tilde C}(t|A,W)=12e^{0.2 W}.
\end{align*}
    \item[B1]$\quad$ The cause specific hazard models and the censoring model are correctly specified, but the propensity score model is not. \begin{align*}
&\lambda_1(t|A=a,W=w)=\lambda_{10}(t)e^{\beta_A^1a+\beta_W^1w},\; 
\lambda_2(t|A=a,W=w)=\lambda_{20}(t)e^{\beta_A^2a+\beta_W^2w}\\
&P(A=1|W)=0.7L+0.1(1-L),\; \lambda_{\tilde C}(t|A,W)=12.
\end{align*}
 \item[B2]$\quad$ The cause specific hazard models  are correctly specified, but the propensity score model and the censoring model are not. 
 \begin{align*}
&\lambda_1(t|A=a,W=w)=\lambda_{10}(t)e^{\beta_A^1a+\beta_W^1w},\; 
\lambda_2(t|A=a,W=w)=\lambda_{20}(t)e^{\beta_A^2a+\beta_W^2w}\\
&P(A=1|W)=0.7L+0.1(1-L),\; \lambda_{\tilde C}(t|A,W)=12e^{0.2 W}.
\end{align*}
    \item[C1]$\quad$  $\lambda_2(t|A=a,W=w)$ is a proportional hazard, but $\lambda_1(t|A=a,W=w)$ is not. The  propensity score model and the censoring model are  correctly specified:
    \begin{align*}
&\lambda_1(t|A=a,W=w)=(1-a)\lambda_{10}(t)e^{\beta_A^1a+\beta_W^1w}+
a\lambda_{10}(t)e^{\beta_A^1L-\beta_A^1(1-L)+\beta_W^1w}
\\
&\lambda_2(t|A=a,W=w)=\lambda_{20}(t)e^{\beta_A^2a+\beta_W^2w}\\
&P(A=1|W)=\mbox{expit}(0+\log{(2)}(W-0.5)),\; \lambda_{\tilde C}(t|A,W)=12.
\end{align*}
 \item[C2]$\quad$  $\lambda_2(t|A=a,W=w)$ is a proportional hazard, but $\lambda_1(t|A=a,W=w)$ is not. The  propensity score model is correctly specified but the censoring model is not:
    \begin{align*}
&\lambda_1(t|A=a,W=w)=(1-a)\lambda_{10}(t)e^{\beta_A^1a+\beta_W^1w}+
a\lambda_{10}(t)e^{\beta_A^1L-\beta_A^1(1-L)+\beta_W^1w}
\\
&\lambda_2(t|A=a,W=w)=\lambda_{20}(t)e^{\beta_A^2a+\beta_W^2w}\\
&P(A=1|W)=\mbox{expit}(0+\log{(2)}(W-0.5)),\; \lambda_{\tilde C}(t|A,W)=12e^{0.2 W}.
\end{align*}
 \nothere{
 \item[D] $\quad$ Neither $\lambda_1(t|A=a,W=w)$  nor $\lambda_2(t|A=a,W=w)$ are  proportional hazards:
     \begin{align*}
&\lambda_1(t|A=a,W=w)=(1-a)\lambda_{10}(t)e^{\beta_A^1a+\beta_W^1w}+
a\lambda_{10}(t)e^{\beta_A^1L-\beta_A^1(1-L)+\beta_W^1w}
\\
&\lambda_2(t|A=a,W=w)=(1-a)\lambda_{20}(t)e^{\beta_A^2a+\beta_W^2w}+a\lambda_{20}(t)e^{\beta_A^1L-
\beta_A^1(1-L)+\beta_W^2w}
\end{align*}
}
\end{itemize}
We used 250 bootstrap replicates to calculate the bootstrap estimate of the variability of $\hat\delta_{1e}(t,1)$. We also  calculated an estimator of the variability based on the squared efficient influence function.
\bigskip

\centerline{Table 2 about here}

\bigskip

The results are summarized in Table 2, where each entry in the table based on 1000 replicates.
We see that both estimators are consistent under scenario A1 and A2, and that the simple estimator is slightly more efficient, as expected since both cause specific hazard functions are proportional. 
In Scenario B1 and B2, where the two cause specific hazards models are correctly specified but the propensity score model is mis-specified, both estimators are consistent.
Under scenario C1, where only    $\lambda_2(t|A,W)$ is a proportional hazard,  the simple estimator is biased whereas the one based on the efficient influence function is still consistent,  as both the censoring and  propensity score models are correctly specified.
Under scenario C2, where the censoring model is mis-specified, the estimator based on the 
efficient influence function is now slightly biased, and the the simple estimator suffer from more severe bias.

\subsection{Prostate cancer data}
\label{sec: prostate cancer example}
%   \subsection*{Prostate cancer application}
To illustrate the new estimators, we used data from a randomized trial on prostate cancer therapy \citep{byar1980choice}, which were also analyzed in
\citet{stensrud2019separable} and are publicly available to anyone  (http://biostat.mc.vanderbilt.edu/DataSets). We
restricted our analysis to the patients who received placebo (127 patients) and high-dose DES (125 patients).  We included baseline measurements of
daily activity function (binary), age (centered around its mean), hemoglobin level (centered around its mean) and previous cardiovascular disease (binary) in our analysis. We considered death due to prostate cancer as the event of interest and death due to other causes (consisting primarily of  cardiovascular deaths) as the competing event. %(5.0 mg of diethylstilbestrol) arm

The events were recorded in monthly intervals from randomization. We used Cox proportional hazards models to obtain the following hazard ratio estimates of  the two cause specific hazards (comparing treatment to placebo): 0.74 (95\% CI: 0.45, 1.21) for the primary cause and     1.17 (95\% CI: 0.82, 1.66) for the competing cause. However, these hazard ratio estimates cannot be interpreted causally \citep{young2018choice, stensrud2019separable}. Yet point estimates of the cumulative incidence curves suggests that the treatment reduces the risk of death death to prostate cancer (Figure 1, left display), but increases the risk of death due to other causes (Figure 1, right display). \bigskip

\centerline{Figure 1 about here}
\bigskip

To disentangle the causal treatment effect on the risk of dying from prostate cancer and from competing events, we therefore estimated
the separable direct $\delta_1(t,0)$ using the proposed $\hat\delta_{1e}(t,0)$
(Figure 2, left display). The point estimates suggest a beneficial separable direct of the prostate cancer therapy (although the confidence bands cover 0).    For example,
the separable direct effect after 40 months is estimated to be approximately -0.09, that is, $\hat\delta_{1e}(t=40,0)=-0.09 \ (95\% \text{ CI: }  -0.17, -0.01 )$, suggesting that a component of treatment reduces the risk of death due to prostate cancer. As discussed in \citet{stensrud2019separable} this is supported by the biological argument that DES  prevents the male testicles from producing testosterone, which, in turn, may prevent prostate cancer cells from replicating. On the other hand, the separable indirect effect (Figure 2, right display) is estimated to be approximately $-0.01 (95\% \text{ CI: }  -0.05, -0.03 )$ at $t=40$, suggesting that there is a component of DES that may increase the risk of death due to other causes, potentially because DES includes estrogen that may increase cardiovascular risk \citep{stensrud2019separable}. Thus, given that our identifiability conditions hold, the point estimates suggests that there is potential for improving the prostate cancer treatment by providing a new (modified) treatment that does not exert effects on death due to other causes. Indeed, such treatments already exist: Luteinising Hormone Releasing Hormone (LHRH) antagonists or orchidectomy (castration) are frequently used to suppress testosterone production in prostate cancer patients today, and these treatments do not contain estrogen. 
\bigskip

\centerline{Figure 2 about here}
\bigskip

\section{Concluding Remarks}
\label{sec: conclusion}
The separable direct and indirect effects clarify the causal interpretation of treatment effects in competing event settings \citep{stensrud2019separable}, which are ubiquitous in medicine and epidemiology. Our new results enable researchers to estimate these effects using classical models statistical models for survival analysis, such as Cox proportional hazards models. Moreover, by deriving the nonparametric efficient influence function,  we have obtained a one-step estimator of the separable effects. Alternatively one may estimate the parameter using Targeted Maximum Likelihood Estimation (TMLE), see \citet{van_der_Laan_Rubin_2006} and \citet{van_der_Laan_2011}. In the Appendix, we give  calculations to carry out TMLE for the specific parameter considered in this paper.

The estimator derived from the efficient influence function has certain desirable robustness properties, allowing some of the working models to be misspecified. It is, however, not straightforward to estimate the variance of the resulting estimator, because it depends on the working models that may contribute to the variability of the estimator. This is a well known problem, see for instance \cite{doi:10.1002/sim.3445} and \cite{van_der_Laan_2011}. In this article, we use the non-parametric bootstrap to estimate the variance. Alternatively, one may follow the more complicated route outlined 
in \cite{10.1093/biomet/asx053} that gives a detailed description of the asymptotic linearity of a similar estimator, although in a simpler setting than the one considered in this paper. Extending the method of \cite{10.1093/biomet/asx053} to the setting considered in the present paper is a topic for future research.

    \bibliographystyle{apalike} % Tells it how you want references displaying in the bibliography.
\clearpage

\bibliography{references}

\clearpage

\section{Appendix}

\subsection{Efficient influence function calculations of the separable direct effects parameter}
We consider the case where we observe the full data $Z=(T,\epsilon, A,W)$, that is, no individual is censored due to loss to follow-up. Once we have the efficient influence function for the full data case then it is easy to get it for the observed data case as described in Section \ref{sec: obs data}.
We calculate the efficient influence function through the  following steps.
\begin{itemize}
\item[(i)]
Calculate the efficient  influence for $\psi(P)=\Lambda_2(t|a,w)$.
\item [(ii)]
Calculate the efficient  influence for 
$$
\psi(P)=Z(t,w)=\frac{\exp{\{-\Lambda_2(t|a,w)\}}}{\exp{\{-\Lambda_2(t|a^{*},w)\}}}.
$$
\item[(iii)]
Calculate the efficient  influence for 
$$
\psi(P)=F_1(t|a^{*},w).
$$
\item[(iv)]
Calculate the efficient  influence for 
$$
\psi(P)=\int_0^t Z(s,w)dF_1(s|a^{*},w).
$$
\item[(v)]
Calculate the efficient  influence for 
$$
\psi(P)=P_1(t,a^{*},a)=\int \int_0^t Z(s,w)dF_1(s|a^{*},w) f(w)dw.
$$
\item[(vi)]
Calculate the efficient  influence for 
$$
\psi(P)=\delta_1(t,1)=P_1(t,a^{*}=1,a=1)-P_1(t,a^{*}=0,a=1),
$$
which is the direct effect we are interested in.
%\item[(vii)] All the above calculations are carried out without  censoring. 
%Finally, calculate the latter one allowing for independent right censoring %adopting formula (10.76) in \cite{tsiatis-book2006}.
\end{itemize}

Since we have posed no structure on the distribution of $(T,\epsilon, A,W)$ we may use the  parametric submodel $P_v$ given by
$$
f_v(x)=f(x)\{1+v\cdot g(x)\},
$$ 
when doing the efficient influence function calculations. In the latter display, $g$ denotes a zero-mean function with finite second moment.
We need to find
$$
\dot{\psi}(g)=\left (\frac{\partial \psi\{P_v\}}{\partial v}\right )_{|_{v=0}},
$$
which we write as
$$
\dot{\psi}(g)=\int \tilde \psi g dP,
$$
and then our efficient influence function is  $\tilde \psi $.

\paragraph{(i)}
We need to write $\Lambda_2(t|a,w)$   first as a function of the underlying probability  measure.
This is done by noting that
$$
P(\epsilon=2|T=s, A=a,w)=\frac{\lambda_2(s| A=a,w)}{\lambda_1(s| A=a,w)+\lambda_2(s| A=a,w)}
$$
and therefore
\begin{equation}
\label{Lam}
\Lambda_{2}(t|a,w)= \int_0^t\frac{f(s,\epsilon=2,a,w)}{P(T>s,a,w)}\, ds
\end{equation}
and, also that 
$$
P(T>s,a,w)=\sum_{\epsilon=1}^2\int_s^{\infty}f(u,\epsilon,a,w)\, du
$$
so  in this way we have  expressed $\Lambda_2(t|a,w)$ in terms  of the underlying density function $f$.
We parametrise the density function
$$
f_v(x)=f(x)\{1+vg(x)\},
$$
and calculate 
$$
\frac{\partial }{\partial v} \left (   \Lambda_{2v}(t|a,w)
   \right )_{|_{v=0}}
$$
where we in \eqref{Lam} replace $f$ with $f_v$.
This leads to 
\begin{align}
\label{derivL}
&\frac{\partial }{\partial v} \left (   \Lambda_{2v}(t|a,w)
   \right )_{|_{v=0}}\notag\\
   =&\int     I(s<t,\epsilon=2)\frac{I(A=a,w)}{P(A=a,w)}\frac{1}{P(T>s|A,w)}\notag\\ 
&\left \{g(s,\epsilon,A,w)-  \frac{\sum_{\tilde\epsilon=1}^2\int_s^{\infty}g(u,\tilde\epsilon,A,w)f(u,\tilde\epsilon,A,w)\, du}{P(T>s,A,w)} \right \}f(s,\epsilon,A,w)dx,
\end{align}
with $x=(s,\epsilon,A,w)$. So we can directly read off that the first term in the latter display contributes with 
$$
 I(T<t,\epsilon=2)\frac{I(A=a,w)}{P(A=a,w)}\exp{\{-\Lambda(T|a,w)\}}=\frac{I(A=a,w)}{P(A=a,w)}\int_0^t\frac{1}{P(T>s|a,w)}dN_2(s)
$$
to the efficient influence  function.  By changing order of integration, it may be seen that second term in \eqref{derivL} contributes
$$
\frac{I(A=a,w)}{P(A=a,w)}\int_0^t\frac{1}{P(T>s|a,w)}I(s\leq T)\lambda_2(s|a,w)\, ds
$$
so all in all, we have that 
\begin{equation}
\label{L2v}
\frac{\partial }{\partial v} \left (   \Lambda_{2v}(t|a,w)
   \right )_{|_{v=0}}=\int \tilde \psi g dP
\end{equation}
where 
\begin{equation}
\label{L2_infl}
\tilde \psi(T,\epsilon,A,w)=\frac{I(A=a,w)}{P(A=a|w)P(w)} \int_0^t\frac{1}{P(T>s|a,w)}dM_2(s|a,w)
\end{equation}
with 
$$
M_2(t|a,w)=N_2(t)-\int_0^tI(s\leq T)\lambda_2(s|a,w)\, ds
$$
being the martingale associated with the cause 2 counting process (where we condition on $A=a$ and $W=w$). 
\paragraph{(ii)}
Taking $h(x,y)=e^{-x+y}$ and by simple differentiation, we get
\begin{align}
\label{S_infl}
\tilde \psi(T,\epsilon,A,w)=
&Z(t,w)\biggl[\frac{I(A=a^{*})}{P(A=a^{*}|w)} \int_0^t\frac{1}{P(T>s|a^{*},w)}dM_2(s|a^{*},w)\notag\\
&-\frac{I(A=a)}{P(A=a|w)} \int_0^t\frac{1}{P(T>s|a,w)}dM_2(s|a,w)\biggr ]\frac{I(W=w)}{P(w)}
\end{align}
\paragraph{(iii)}
We use $L=(A,W^T)^T$.
$$
F_1(t|l)=\int_0^tP(T>s|l)\lambda_1(s|l)\, ds
$$
We seek to write this parameter as a function of the underlying probability measure, $\psi(P)$. The density function is
$$
f(t,\epsilon,l)=P(T>t|l)\{\lambda_1(s|l)\}^{I(\epsilon=1)}\{\lambda_2(s|l)\}^{I(\epsilon=2)} f(l)
$$
and we can therefore write $F_1(t|l)$ as
$$
F_1(t|l)=\int I(s<t,\epsilon=1)\frac{I(L=l)}{P(L=l)}f(s,\epsilon,l)d(s,\epsilon,l)=\psi(P)
$$
We now parametrise the density function
$$
f_v(x)=f(x)\{1+vg(x)\},
$$
where $x=(s,\epsilon,l)$. 
So we first need to find $\left (\frac{\partial \psi(P_v)}{\partial v}\right )_{|_{v=0}}$, which essentially amounts to find
$$
\frac{\partial }{\partial v} \left (   \frac{f_v(s,\epsilon,l)}{P_v(L=l)}
   \right )_{|_{v=0}}
$$
It is easily seen that
\begin{align*}
\frac{\partial }{\partial v} \left (   P_v(L=l)
   \right )_{|_{v=0}}&=E\{g(X)|L=l\}P(L=l)\\
  \frac{\partial }{\partial v} \left (   f_v(s,\epsilon,l)
   \right )_{|_{v=0}} &=g(x)f(x)
\end{align*}
giving that
\begin{align*}
\left (\frac{\partial \psi(P_v)}{\partial v}\right )_{|_{v=0}}=\int I(s<t,\epsilon=1)\frac{I(L=l)}{P(L=l)}\{g(x)-E(g(x)|L=l)\} f(x)\, dx.
\end{align*}
Further,
\begin{align*}
\int I(s<t,\epsilon=1)\frac{I(L=l)}{P(L=l)}&\{E(g(x)|L=l)\} f(x)\, dx=E\{I(T<t,\epsilon=1)\frac{I(L=l)}{P(L=l)}E(g(x)|L) \}\\
&=EE(\mbox{   }|L)\\
&=E\left (g(X)\frac{I(L=l)}{P(L=l)}E\{I(T<t,\epsilon=1)|L\}\right )
\end{align*}
giving us that 
\begin{align*}
\left (\frac{\partial \psi(P_v)}{\partial v}\right )_{|_{v=0}}&=\int g(x)\frac{I(L=l)}{P(L=l)} \left [ I(s<t,\epsilon=1)-E\{I(T<t,\epsilon=1)|L=l\}\right ] f(x)\, dx\\
&=\int \tilde \psi g dP,
\end{align*}
where 
\begin{align}
\label{F1_infl}
\tilde \psi(T,\epsilon,A=a^{*},w)=
\frac{I(A=a^{*},w)}{P(A=a^{*}|w)P(w)} \{N_1(t)-F_1(t|a^{*},w)\},
\end{align}
is the efficient influence function. 

\paragraph{(iv)}
By looking at the transform
$$
\int x(s)dy(s)
$$
we obtain that the efficient influence function of  $
\psi(P)=\int_0^t Z(s,w)dF_1(s|a^{*},w)
$ consists of the sum of the terms in the following two displays
\begin{align}
\label{1_infl}
&\biggl[\frac{I(A=a^{*})}{P(A=a^{*}|w)} \int_0^t \int_u^tZ(s,w)dF_1(s|a^{*},w)\frac{1}{P(T>u|a^{*},w)}dM_2(u|a^{*},w)\notag\\
&-\frac{I(A=a)}{P(A=a|w)} \int_0^t\int_u^tZ(s,w)dF_1(s|a^{*},w)\frac{1}{P(T>u|a,w)}dM_2(u|a,w)\biggr ]\frac{I(W=w)}{P(w)}
\end{align}
and 
\begin{align}
\label{2_infl}
\frac{I(A=a^{*},w)}{P(A=a^{*}|w)P(w)} \int_0^tZ(s,w)\{dN_1(s)-dF_1(s|a^{*},w)\}.
\end{align}
\paragraph{(v)}
We  obtain directly that the efficient influence function of  
$$
\psi(P)=\int \int_0^t Z(s,w)dF_1(s|a^{*},w)f(w) dw
$$
 is
\begin{align}
\label{E_infl}
&\int_0^t  \int_u^tZ(s,w)dF_1(s|a^{*},w) \biggl [\frac{dM_2(u|a^{*},w)}{P(T>u|a^{*},w)}\frac{I(A=a^{*})}{P(A=a^{*}|w)} 
- \frac{dM_2(u|a,w)}{P(T>u|a,w)}\frac{I(A=a)}{P(A=a|w)}\biggr ]\notag \\
&-\frac{I(A=a^{*})}{P(A=a^{*}|w)} \int_0^tZ(s,w)\{dN_1(s)-dF_1(s|a^{*},w)\}\notag \\
&+ \int_0^tZ(s,w)dF_1(s|a^{*},w)-E\biggl \{\int_0^tZ(s,W)dF_1(s|a^{*},W)\biggr \}
\end{align}
Note that 
$$
\int_u^tZ(s,w)dF_1(s|a^{*},w) =P_1(t,a^{*},a,w)-P_1(u,a^{*},a,w).
$$
\paragraph{(vi)}
We can now use the result in (v) to get the efficient influence function of
$$
\psi(P)=\delta_1(t,1)=P_1(t,a^{*}=1,a=1)-P_1(t,a^{*}=0,a=1)
$$
since both terms on the right hand side of the latter display are special cases of (v). This gives the  desired efficient influence function $\tilde \psi(t,T,\epsilon,A,W)$
\begin{align}
\label{Del1_infl}
&\bigl\{ N_1(t)-P_1(t,1,1,W)\bigr \}\frac{I(A=1)}{P(A=1|W)}    
-\bigl \{ \int_0^t\frac{e^{-\Lambda_2(s|1,W)}}{e^{-\Lambda_2(s|0,W)}}dN_1(s)-P_1(t,0,1,W)\bigr \} \frac{I(A=0)}{P(A=0|W)}
\notag\\
&-\int_0^t   \bigl \{P_1(t,0,1,W)-P_1(u,0,1,W)\bigr \} \biggl [\frac{dM_2(u|0,W)}{P(T>u|0,W)}\frac{I(A=0)}{P(A=0|W)} 
- \frac{dM_2(u|1,W)}{P(T>u|1,W)}\frac{I(A=1)}{P(A=1|W)}\biggr ] \notag \\
&+\delta_1(t,1,W)- E\bigl\{ \delta_1(t,1,W)\bigr \}
\end{align}
where, as earlier defined,
$$
\delta_1(t,1,W)=P_1(t,1,1,W)-P_1(t,0,1,W).
$$
Note also that 
$$
P_1(t,1,1,W)=F_1(t|A=1,W),
$$
the cumulative incidence function for cause 1 given $A=1$ and $W$, and that 
$$
\delta_1(t,1)= E\bigl\{ \delta_1(t,1,W)\bigr \}
$$
is the parameter of interest. 

The  efficient influence function  $\tilde \psi(t,T,\epsilon,A,W)$ needs to be an element of 
the tangent space
$\mathcal{T}$, which is given by

$$\mathcal{T}=\mathcal{T}_1\oplus\mathcal{T}_2\oplus\mathcal{T}_3$$
where 
\begin{align*}
\mathcal{T}_1&=\bigl \{ \int h(u,A,W)dM_1(u,A,W)\mbox{ for all $h(u,a,w)$}\bigr \},\\
\mathcal{T}_2&=\bigl \{ \int h(u,A,W)dM_2(u,A,W)\mbox{ for all $h(u,a,w)$}\bigr \},\\
\mathcal{T}_3&=\bigl \{h(A,W)\in \mathcal{H}: E\{h(X,W)\}=0\bigr \}.
\end{align*}
This is not obvious at first sight when looking at \eqref{Del1_infl}. However, it turns out be true as we may show that
\begin{equation}
\label{simple_identity}
N_1(t)-F_1(t|a,w)=\int_0^tdM_1(s|a,w)-\int_0^t \frac{\{F_1(t|a,w)-F_1(s,|a,w)\bigr \}}{P(T>s|a,w)}dM(s|a,w)    
\end{equation}
where $M(t|a,w)=M_1(t|a,w)+M_2(t|a,w)$. To see that \eqref{simple_identity} holds just check that it holds in the two cases $T\leq t$ and $t<T$. Using  \eqref{simple_identity} we can write the   efficient influence function $\tilde \psi(t,T,\epsilon,A,W)$ as 
\begin{align}
\label{Del1_infl_expr2}
&\bigl\{ M_1(t|1,W)- \int_0^t \frac{\{F_1(t|1,w)-F_1(s,|1,w)\bigr \}}{P(T>s|1,w)}dM(s|1,w)  \bigr \}\frac{I(A=1)}{P(A=1|W)}\notag   \\ 
&-\bigl \{ \int_0^t\frac{e^{-\Lambda_2(s|1,W)}}{e^{-\Lambda_2(s|0,W)}}\bigl\{ dM_1(s|0,w)- \frac{\{F_1(t|0,w)-F_1(s,|0,w)\bigr \}}{P(T>s|0,w)}dM(s|0,w)\bigr\}
 \frac{I(A=0)}{P(A=0|W)}
\notag\\
&+\int_0^t   \bigl \{P_1(t,0,1,W)-P_1(u,0,1,W)\bigr \} \biggl [ \frac{dM_2(u|1,W)}{P(T>u|1,W)}\frac{I(A=1)}{P(A=1|W)}-\frac{dM_2(u|0,W)}{P(T>u|0,W)}\frac{I(A=0)}{P(A=0|W)} 
\biggr ] \notag \\
&+\delta_1(t,1,W)- E\bigl\{ \delta_1(t,1,W)\bigr \}
\end{align}
and from this representation we note that  $\tilde \psi(t,T,\epsilon,A,W)$  indeed belongs to the tangent space
$\mathcal{T}$. The expression in the latter display can further be re-written giving
\eqref{Del1_infl_expr3}.

\subsection{Robustness properties}
\label{sec: robustness App}
  We argue now that the estimator $\hat \delta_{1e}(t,1)$ given in
\eqref{EIF-est} has certain robustness properties unlike the initial estimator $\hat \delta_1(t,1)$ based on Cox-models for the cause specific hazard functions.
We first consider the case where there is no censoring.
The estimator $\hat \delta_{1e}(t,1)$ consist of a difference between two terms, and we consider these terms separately. We have that
$
P_1(t,1,1)=E\{F_1(t|A=1,W)\},
$ and define 
$$Y=\frac{N_1(t)I(A=1)}{P(A=1|W)}
$$
so 
$$
E(Y)=E\{F_1(t|A=1,W)\}
$$
if $P(A=1|W)$ is correctly specified.
The corresponding terms of the estimator are
\begin{align}
\label{DR1}
\biggl [ N_1(t)-F_1(t|A=1,W)\biggr ]&\frac{I(A=1)}{P(A=1|W)}    +F_1(t|A=1,W)\notag\\
&=Y+F_1(t|A=1,W)\left\{ \frac{P(A=1|W)-I(A=1)}{P(A=1|W)}\right\}
\end{align}
from which we see that the mean of the left hand side of \eqref{DR1}, if $P(A=1|W)$ is correctly specified, is
$$
E(Y)+0=F_1(t,1)
$$
so it has the correct mean in that case. On the other hand, if $F_1(t|A=1,W)$ is correctly specified (which is the case if $\lambda_j(t|A=1,W)$, $j=1,2$, are correctly specified)
but  $P(A=1|W)$  may not be so, then the mean of 
 the left hand side of \eqref{DR1} is 
 $$
 E\left \{ \frac{F_1(t|A=1,W)P_0(A=1|W)}{ P^*(A=1|W)}\right \}+ E\left [F_1(t|A=1,W) \left \{1-\frac{P_0(A=1|W)}{ P^*(A=1|W)}\right\}\right ]=F_1(t,1),
 $$
 where we use $ P^*(A=1|W)$ to denote the limit (in probability) of a potentially mis-specified estimator  $\hat P(A=1|W)$, and $P_0$ to denote the truth.
 So the estimator
 $$
 \hat P(t,1,1)=n^{-1}\sum_i \biggl[  \hat P(t,1,1,W_i)+\bigl\{ N_{i1}(t)-\hat P_1(t,1,1,W_i)\bigr \}\frac{I(A_i=1)}{\hat P(A=1|W_i)}  \biggr ]
 $$
 is double robust in this sense. We now turn to the more difficult part, $ \hat P(t,0,1)$. We need to look at
 \begin{align}
\label{DR2_term}
&\bigl \{ \int_0^t\frac{e^{-\Lambda_2(s|1,W)}}{e^{-\Lambda_2(s|0,W)}}dN_1(s)-P_1(t,0,1,W)\bigr \} \frac{I(A=0)}{P(A=0|W)}+P_1(t,0,1,W)
\notag\\
&+\int_0^t   \bigl \{P_1(t,0,1,W)-P_1(u,0,1,W)\bigr \} \biggl [\frac{dM_2(u|0,W)}{P(T>u|0,W)}\frac{I(A=0)}{P(A=0|W)} 
-\frac{dM_2(u|1,W)}{P(T>u|1,W)}\frac{I(A=1)}{P(A=1|W)}\biggr ] .
\end{align}
Now let
$$
Y=\int_0^t\frac{e^{-\Lambda_{20}(s|1,W)}}{e^{-\Lambda_{20}(s|0,W)}}dN_1(s)\frac{I(A=0)}{P(A=0|W)}
$$
where $\Lambda_{20}$ denotes the true value of the parameter. If $P(A=1|W)$ is correctly specified then
$$
E(Y)=P_1(t,0,1).
$$
Display \eqref{DR2_term} can be rewritten as
 \begin{align}
\label{DR2_term_rewr}
&Y+P_1(t,0,1,W)\left\{ \frac{P(A=0|W)-I(A=0)}{P(A=0|W)}\right\}\notag \\
&+\int_0^t\bigl \{ \frac{e^{-\Lambda_2(s|1,W)}}{e^{-\Lambda_2(s|0,W)}}-\frac{e^{-\Lambda_{20}(s|1,W)}}{e^{-\Lambda_{20}(s|0,W)}}\bigr \} dN_1(s)\frac{I(A=0)}{P(A=0|W)}
\notag\\
&+\int_0^t Z(s,W)
\biggl [\int_0^s\frac{dM_2(u|0,W)}{P(T>u|0,W)}\frac{I(A=0)}{P(A=0|W)} 
-\int_0^s\frac{dM_2(u|1,W)}{P(T>u|1,W)}\frac{I(A=1)}{P(A=1|W)}\biggr ] dF_1(s|0,W).
\end{align}
Assume first that $\lambda_1(t| j,W)$ and $\lambda_2(t| j,W)$, $j=0,1$, are correctly specified but  $P(A=j|W)$ may not be so.
It is then clear from \eqref{Del1_infl_expr3} that the estimator based on the efficient influence function is unbiased as the efficient influence function will still have zero mean.

%The second term of display \eqref{DR2_term_rewr} disappears, and the mean of the third term of display %\eqref{DR2_term_rewr} is zero as $M_2$ is then a martingale conditioned on $W$. The mean of the first term is 
% $$
% E\left \{ \frac{P_1(t,0,1,W)P_0(A=0|W)}{\tilde P(A=0|W)}\right \}+ E\left [P_1(t,0,1,W) \left %\{1-\frac{P_0(A=0|W)}{\tilde P(A=0|W)}\right\}\right ]=P_1(t,0,1).
% $$
 We now assume that  $P(A=j|W)$ and $\lambda_2(t| j,W)$  are 
 correctly specified but   $\lambda_1(t| j,W)$  may be mis-specified. It follows directly that
 the mean of the first term of  display \eqref{DR2_term_rewr} is
 $$
 E(Y)+0=P_1(t,0,1).
 $$
 and it is also clear that the  the mean of the sum of the last two terms of  display \eqref{DR2_term_rewr} is zero as long as $\lambda_2(t| j,W)$ is correctly specified.
 
  We now assume that  $P(A=j|W)$ and $\lambda_1(t| j,W)$  are 
 correctly specified but   $\lambda_2(t| j,W)$  may be mis-specified. It follows again that
 the mean of the first term of  display \eqref{DR2_term_rewr} is
 $$
 E(Y)+0=P_1(t,0,1).
 $$
 Keep in mind that we have 
 \begin{align}
\label{dN1}
 E\left \{ dN_1(s)\frac{I(A=0)}{P(A=0|W)} |W
\right \}=dF_{10}(s|0,W)
\end{align}
 and that
  $$
 E\left \{ dN_2(s)\frac{I(A=j)}{P(A=j|W)} |W
\right \}=dF_{20}(s|j,W)
 $$
 for $j=1,2$.
 We let $\Lambda^{*}_k(t|j,w)$ denote the limit of a given estimator $\hat \Lambda_k(t|j,w)$ for $k=1,2$ and $j=0,1$, and similarly, let $Z^{*}(t,w)$ denote the limit of the estimator $\hat Z(t,w)$, and $ P^*(t>u|j,w)$ being $e^{-\Lambda^{*}_{2}(u|j,w)-\Lambda^{*}_{2}(u|j,w)}$ (all supposed to exist such as would be the case if we use Cox-models for the two cause specific hazard functions).
 We note that 
 \begin{align}
 \label{ThirdTerm}
 E\bigl \{ \frac{dN_2(u)-Y(u)d\Lambda^{*}_2(u|j,W)}{ P^*(T>u|j,W)}\frac{I(A=j)}{P(A=j|W)} |W\bigr\} 
 &=\frac{dF_{20}(u|j,W))}{ P^*(T>u|j,W)}-\frac{P_0(T>u|j,W)d \Lambda_{2}^{*}(u|j,W)}{ P^*(T>u|j,W)}\notag \\
 &= -\frac{P_0(T>u|j,W)}{P^*(T>u|j,W)}\bigl\{ d\Lambda^{*}_{2}(u|j,W)- \Lambda_{20}(u|j,W)\bigr\}\notag \\
 &=-e^{\{\Lambda^{*}_{2}(u|j,W)- \Lambda_{20}(u|j,W)\}}\bigl\{ d\Lambda^{*}_{2}(u|j,W)- \Lambda_{20}(u|j,W)\bigr\}
 \end{align}
 since $\lambda_1$ is correctly specified. Integrating the right hand side of \eqref{ThirdTerm} from 0 to $s$ gives
 $$
 -e^{\{\Lambda^{*}_{2}(u|j,W)- \Lambda_{20}(u|j,W)\}}+1
 $$
%\textcolor{red}{ OBS, OBS the following is not yet complete, need some more details...}
 %Also,
 %$$
 %\frac{P_0(T>u|j,W)}{\tilde P(T>u|j,W)}dF_1(s|0,W).
 %$$
 Using that
 $$
  dF_1(s|0,W)=e^{-\{\Lambda^{*}_{2}(u|0,W)- \Lambda_{20}(u|0,W)\}}  dF_{10}(s|0,W)
 $$
 we see that the last term of \eqref{DR2_term_rewr} converges to
 $$
 -\int_0^t\{ Z^{*}(s,w)-Z_0(s,w)\} dF_{10}(s|0,W)
 $$
 that cancels with the limit of the second term of \eqref{DR2_term_rewr}.
 Hence, $\hat \delta_{1e}(t,1)$ is consistent if two of the three models
(i): $\lambda_1(t|j,w)$; (ii): $\lambda_2(t|j,w)$  and (iii): $P(A=j|W)$, are correctly specified. We now consider the case with censoring and let (iv): $\lambda_C(s|A,W)$ be the censoring model.
Using \eqref{Tsiatis-fomula}, we can write 
\begin{align}
\label{Tsiatis-fomula2}
   \psi(t,D)&=\tilde \psi(t,Z)+\int_0^{\infty}\frac{\bigl[\tilde \psi(t,Z)-E\bigl\{ \tilde \psi(t,Z) |T>s, A,W \bigr\}\bigr]}{K_C(s|A,W)}dM_C(s|A,W)
\end{align}
that depends on $G$ as noted in Section \ref{sec: robustness}.
We let $G^*$ denote the limit of the estimator $G_n$ of $G$, and have, as noted above, that  $E\{\tilde \psi(t,Z,G^*)\}=0$ if two out of the three models (i)-(iii) are correctly specified.
Keep in mind that $dM_C(s|A,W)=dN_C(s)-I(s\le \tilde T)d\Lambda_C(s|A,W)$. 
Since $T$ and $C$ are conditional independent given $A$ and $W$, we have
\begin{align*}
 E\{dN_C(s)|T,A,W)\}&=I(s\leq T)P(T>s|A,W)K_C(s|A,W)\lambda_C(s|A,W)\\
  E\{I(s\le \tilde T)|T,A,W)\}&=I(s\leq T)P(T>s|A,W)K_C(s|A,W).
\end{align*}
Therefore,
\begin{align*}
&E\biggl\{
\frac{\bigl[\tilde \psi(t,Z,G^*)-E\bigl\{ \tilde \psi(t,Z) |T>s, A,W,G^* \bigr\}\bigr]}{K^*_C(s|A,W)}dM^{*}_C(s|A,W)\biggr \}\\
=
&E\biggl\{
I(s\leq T)\bigl[\tilde \psi(t,Z,G^*)-E\bigl\{ \tilde \psi(t,Z) |T>s, A,W,G^* \bigr\}\bigr]\frac{K_C(s|A,W)}{K^*_C(s|A,W)}
\{\lambda_C(s|A,W)-\lambda^*_C(s|A,W)\}\, ds\biggr \},
\end{align*}
and 
\begin{align*}
&E\bigl\{
I(s\leq T)\bigl[\tilde \psi(t,Z,G^*)-E\bigl\{ \tilde \psi(t,Z) |T>s, A,W,G^* \bigr\}\bigr]|A,W\bigr \}\\
=
&\sum_{\epsilon}\int_s^{\infty}\tilde \psi(t,u,\epsilon,A,W)
\bigl[f(u,\epsilon|A,W)-f^{*}(u,\epsilon|A,W)\frac{P(T>u|A,W)}{P^*(T>u|A,W)}\bigr ]\, du
\end{align*}
We hence see that the mean of the latter term in \eqref{Tsiatis-fomula2} is zero if either 
models (i) and (ii) are correctly specified or if model (iv) is correctly specified.
The proposed estimator $\hat \delta_{1e}(t,1)$ is thus consistent if the following models are correctly specified: (i) and (ii), or (i), (iii) and (iv), or (ii), (iii) and (iv).

\subsection{Large sample properties of the simple estimator}
\label{App: large sample properties}

%\subsubsection{The simple estimator}
%\label{App: large sample properties, simple estimator}

It follows that under appropriate conditions, as in \citet{10.2307/25734118}, that 
\begin{align}
\label{asym_decomp}
n^{1/2}\left\{\hat P_1(t,a_Y,a_D)- P_1(t,a_Y,a_D)\right \}=
&n^{-1/2}\sum_i\left[ P_1(t,a_Y,a_D,W_i)- E \left \{P_1(t,a_Y,a_D,W_i)\right \}\right ]\nonumber\\
&+E\left [ n^{1/2}\left\{\hat P_1(t,a_Y,a_D,W)- P_1(t,a_Y,a_D,W)\right \} \right ]+o_p(1),
\end{align}
where the expectation in the second term is taken wrt $W$. The first term on the right hand side of
\eqref{asym_decomp} is already on iid-form, so we only need to deal with the second term.
Define 
$$
\theta_t(a_Y,a_D, w)=e^{-\Lambda_{1}(t|A=a_Y,w)-\Lambda_{2}(t|A=a_D,w)}.
$$
It is then easily seen that 
\begin{align}
\label{decomp}
n^{1/2}\left\{d\hat P_1(t,a_Y,a_D,W)- dP_1(t,a_Y,a_D,W)\right \} =
&n^{1/2}\left\{ \hat\theta_t(a_Y,a_D, W)-\theta_t(a_Y,a_D, W)\right\}e^{ \beta_A^1 a_Y+\beta_W^1 W}d\Lambda_{10}(t)\nonumber\\
&+ \theta_t(a_Y,a_D, W)g(\beta^1,a_Y,W)d\, n^{1/2}\left\{ \hat\Lambda_{10}(t)-\Lambda_{10}(t)\right\}\\
&+ \theta_t(a_Y,a_D, W)\left\{D_{\beta^1}g(\beta^1,a_Y,W)\right\} n^{1/2}\left\{ \hat\beta^1-\beta^1\right \}d\Lambda_{10}(t)\nonumber\,
\end{align}
and we then need to find the influence functions of the three terms on the right hand side of \eqref{decomp}, and take expectation wrt $W$. 
In the latter display,
$$
g(\beta^1,a_Y,W)=e^{ \beta_A^1 a_Y+\beta_W^1 W}
$$
with $\beta^j=(\beta_A^j,\beta_W^j)$, $j=1,2$.
The second and third terms can be dealt with directly. For instance, with $\epsilon_i^{\beta^1}$ being the influence function corresponding to $ \hat\beta^1$, 
\begin{align*}
E_W\bigl [\theta_t(a_Y,a_D, W)\left\{D_{\beta^1}g(\beta^1,a_Y,W)\right\} &n^{1/2}\left\{ \hat\beta^1-\beta^1\right \}d\Lambda_{10}(t)\bigr ]=\\
&E_L\bigl [\theta_t(a_Y,a_D, L)\left\{D_{\beta^1}g(\beta^1,a_Y,L)\right\} \bigr ]\left\{n^{-1/2}\sum_i\epsilon_i^{\beta^1}\right\}d\Lambda_{10}(t)
\end{align*}
in which we may estimate $E_W\bigl [\theta_t(a_Y,a_D, W)\left\{D_{\beta^1}g(\beta^1,a_Y,W)\right\} \bigr ]$ by
$$
n^{-1}\sum_i\theta_t(a_Y,a_D, W_i)\left\{D_{\beta^1}g(\beta^1,a_Y,W_i)\right\}. 
$$
and then replacing parameters with their estimates.
Similarly with the second term on the right hand side of \eqref{decomp}. 
The first term on the right hand side of \eqref{decomp} is handled using the mean value theorem (Taylor expansion) with respect to the parameters $\{\Lambda_0^j(t),\beta^j\}$, $j=1,2$. The expectation wrt $W$ of the part corresponding to $\lambda_{10}(t)$ of this term is 
$$
-E_W\bigl [\theta_t(a_Y,a_D, W)e^{ 2(\beta_A^1 a_Y+\beta_W^1 W)}\bigr ]\left\{n^{-1/2}\sum_i\epsilon_i^{\lambda_{10}}(t)\right\}d\Lambda_{10}(t),
$$
where $E_W\bigl [\theta_t(a_Y,a_D, W)e^{ 2(\beta_A^1 a_Y+\beta_W^1 W)}\bigr ]$ is estimated by
$$
n^{-1}\sum_i \theta_t(a_Y,a_D, W_i)e^{ 2(\beta_A^1 a_Y+\hat\beta_W^1 W_i)}
$$
and replacing parameters with their estimates. Similarly with the other terms.
By combining all these terms, we have that
$$
n^{1/2}\left\{\hat P_1(t,a_Y,a_D)- P_1(t,a_Y,a_D)\right \}=\sum_i \epsilon_i^{P_1}(t,a_Y,a_D),
$$
where  	$\epsilon_i^{P_1}(t,a_Y,a_D)$ are zero-mean iid terms (i.e.\ we have found the influence function), and we have also a way of estimating the influence function. So, basically, what we need is the first term in \eqref{asym_decomp}, and then all the influence functions corresponding to  the parameters $\{\Lambda_{j0}(t),\beta^j\}$ using the usual Cox-regression estimators (these can be extracted from the {\tt coxaalen}-function in R-package {\tt timereg}), and then also some ordinary empirical means such as
$$
n^{-1}\sum_i \theta_t(a_Y,a_D, W_i)e^{ 2(\beta_A^1 a_Y+\hat\beta_W^1 W_i)}.
$$

\subsection{Targeted Maximum Likelihood Estimation}\label{sec: TMLE}
\noindent We here briefly 
describe how to do Targeted Maximum Likelihood Estimation (TMLE), see
\cite{van_der_Laan_Rubin_2006} and \cite{van_der_Laan_2011}. We do so for the setting with no censoring,  based on expression \eqref{Del1_infl_expr3}. The generalization to the censored case follows immediately.
The tangent space adhering to $f(A|W)$ consist of all functions $r(A,W)$ so that $E\{r(A,W)|W\}=0$, and the efficient influence function $\tilde \psi(t,T,\epsilon,A,W)$ is therefore orthogonal to that space. To see why,
$$
E\bigl \{r(A,W)\int g(s;A,W)dM_j(s|A,W)\bigr \}=E\bigl [r(A,W)E\{\int g(s;A,W)dM_j(s|A,W)|A,W\}\bigr ]=0
$$
and 
$$
E\bigl [r(A,W)\{ \delta_1(t,1,W)- \delta_1(t,1)\} \bigr ]=E\bigl [ \{ \delta_1(t,1,W)- \delta_1(t,1)\}E\{r(A,W)|W\} \bigr ]=0.
$$
Hence we need not to fluctuate the initial estimator of $P(A=a|W)$, and can concentrate on the part of the tangent space adhering to
 $f(T,\epsilon|A,W)$.
The needed parametric submodels are obtained by parametrizing the cause specific hazard function as 
$$
\lambda_{j,\gamma}(s|A,W)=\lambda_{j}(s|A,W)e^{\gamma H_j(s;A,W)}, \quad j=1,2.
$$ 
The score (in $\gamma$) corresponding to  $f(T,\epsilon|A,W)$ of the likelihood  is 
$$
\sum_{j=1}^2\int H_j(s;A,W)dM^T_j(s|A,W)
$$
so we see from  \eqref{Del1_infl_expr3} that we need to pick the functions $H_j(s;A,W)$ as 
\begin{align*}
H_1(s;A,W)=&I(s\leq t)g(A,W)\frac{e^{-\Lambda_2(s|1,W)}}{e^{-\Lambda_2(s|A,W)}}\biggl\{ 1-  \frac{\{F_1(t|A,w)-F_1(s,|A,w)\bigr \}}{P(T>s|A,w)} \biggr\}\\
H_2(s;A,W)=&I(s\leq t)\frac{g(A,W)}{P(T>s|A,W)}, \biggl \{P_1(t,0,1,W)-P_1(s,0,1,W)-\\
&\frac{e^{-\Lambda_2(s|1,W)}}{e^{-\Lambda_2(s|A,W)}}\bigl \{F_1(t|A,w)-F_1(s,|A,W)\bigr \} \biggr \}. \\
\end{align*}
We find $\hat \gamma$ as the solution to 
\begin{equation}
\label{TMLE}
0=U(\gamma)=\sum_{i=}^n\sum_{j=1}^2\int H_j(s;A_i,W_i)\bigl \{ dN_{ij}(s)-Y_i(s)e^{\gamma H_j(s;A_i,W_i)}d\hat\Lambda_j^{0}(s|A_i,W_i)\bigr\},
\end{equation}
where
$\hat \Lambda_j^{0}(s|A_i,W_i)$ is an initial estimate of $\Lambda_j(s|A_i,W_i)$ based for instance on semiparametric models such as the Cox-model.
We then calculate 
$$
\hat \Lambda_j^{(1)}(s|A,W)=\int_0^s e^{\hat \gamma H_j(u;A,W)}d\hat\Lambda_j^{0}(s|A,W).
$$
Next step is to  solve equation \eqref{TMLE} with $\hat \Lambda_j^{0}$ replaced by $\hat \Lambda_j^{(1)}$ to get an updated $\hat \gamma^{(1)}$ to calculate
$$
\hat \Lambda_j^{(2)}(s|A,W)=\int_0^s e^{\hat \gamma^{(1)} H_j(u;A,W)}d\hat\Lambda_j^{1}(s|A,W)
$$
and iterate until convergence (ie until $\hat\gamma$  does no longer change), say it happens at iteration step $k$. The TMLE estimate of  parameter of interest is then
simply the plug-in estimate
\begin{align*}
\hat\delta_1(t,1)=
 &n^{-1}\sum_i\left \{\int_0^te^{-\hat\Lambda^{(k)}_1(s|1,W_i)-\hat\Lambda^{(k)}_2(s|1,W_i)}d\hat\Lambda^{(k)}_1(s|1,W_i)\right\} \\
 - &n^{-1}\sum_i\left \{\int_0^te^{-\hat\Lambda^{(k)}_1(s|0,W_i)-\hat\Lambda^{(k)}_2(s|1,W_i)}d\hat\Lambda^{(k)}_2(s|0,W_i)\right\}.
\end{align*}
\bigskip
\bigskip
\bigskip
\bigskip
\bigskip

   \begin{table}[ht!]
\caption{Simulation results concerning the estimators $\hat P_1(t,1,1)$, $\hat P_1(t,0,1)$ and $\hat \delta_1(t,1)=\hat P_1(t,1,1)-\hat P_1(t,0,1)$. Each entry in the table is based on 1000 replicates.
 \bigskip}

{\small
\begin{center}
\begin{tabular}{|l|ccc|ccc|}
\hline
 &\multicolumn{3}{c|}{n=400 }&\multicolumn{3}{c|}{ n=800 } \\
 & $t=2$ & $t=4$ & $t=6$ & $t=2$ & $t=4$ & $t=6$\\

\hline
%400&& \multicolumn{4}{c|}{ $F=40, K=5$}&\multicolumn{4}{c|}{ $F=126, K=0$}\\
%&Bias $\hat \beta_X$ &0.001 &-& -& -&0.001 &-& -& -\\
%&95\% CP($\hat \beta_X$)&97.7 &- & - & -&94.9 &- & - & -\\
%800&& \multicolumn{4}{c|}{ $F=79, K=0$}&\multicolumn{4}{c|}{ $F=249, K=0$}\\
True $ P_1(t,1,1)$ &  0.052        &0.089  &       0.118 &     0.052        &0.089  &       0.118   \\
Mean $ \hat P_1(t,1,1)$  &0.052 &0.089 &0.118&   0.052&0.089 &       0.118     \\
sd ($ \hat P_1(t,1,1)$)&0.013 &0.020 &0.025 & 0.009&0.014 &0.018 \\
see ($ \hat P_1(t,1,1)$)&0.013&0.020 &0.025 &0.009  &       0.014&0.018   \\
95\% CP($ \hat P_ 1(t,1,1)$)&0.924& 0.934& 0.945  &   0.939          & 0.938 &     0.942 \\
\hline
True $ P_1(t,0,1)$  & 0.100             &0.170        & 0.219    &    0.100             &0.170         & 0.219       \\
Mean $ \hat P_1(t,0,1)$  &0.099 &0.169& 0.218 &     0.100    &     0.170 &        0.219     \\
sd ($ \hat P_1(t,0,1)$)&0.020 & 0.027 &0.032&0.0143         &  0.019   & 0.023 \\
see ($ \hat P_1(t,0,1)$)&0.019 &0.028 &0.033&0.014        & 0.019 & 0.023 \\
95\% CP($ P_1(t,0,1)$)&0.929&0.952&0.952   &    0.948          & 0.954 &      0.956 \\
\hline
True $  \delta_1(t,1)$  &  -0.049        &  -0.080 & -0.101  &       -0.049        &  -0.080 & -0.101         \\
Mean $ \hat \delta_1(t,1)$  &-0.048&-0.079 &-0.099&      -0.049 &-0.081 &-0.101     \\
sd ($  \hat\delta_1(t,1)$)&0.019&0.031&0.038 &  0.014  & 0.022 & 0.027 \\
see ($ \hat \delta_1(t,1)$)&0.019&0.031&0.039& 0.014& 0.022& 0.028 \\
95\% CP($  \hat\delta_1(t,1)$)&0.946&0.951&0.952   &   0.956           &  0.957&       0.956\\
\hline
%Bias $\hat B_{1}(t)$ &3200&0.002   &   0.000     &  -0.001&2000&       0.000&-0.002   &  - 0.006      \\
%sd ($\hat B_{1}(t)$)&&0.043 &0.090& 0.154& &0.028&0.061 &0.101\\
%see ($\hat B_{1}(t)$)&&0.043 &0.091& 0.150& &0.029&0.062 &0.103\\
%95\% CP($\hat B_{1}(t)$)&&95.2&     95.7&      95.2   &&   95.5 &95.4&     95.8\\
%Bias $\hat B_{2}(t)$ &&-0.003   &   -0.003     &  -0.002& &       0.001&0.002   &  - 0.002      \\
%sd ($\hat B_{2}(t)$)&&0.053 &0.118& 0.194& &0.038&0.082 &0.136\\
%see ($\hat B_{2}(t)$)&&0.053 &0.115& 0.194& &0.038&0.082 &0.138\\
%95\% CP($\hat B_{2}(t)$)&&95.6&     94.8&      95.9   &&   95.6 &95.5&     95.9\\
%Bias $\tilde B_{1}(t)$ &&-0.033&-0.100   &  - 0.165& &      -0.033&-0.100   &  - 0.167      \\
%Bias $\tilde B_{2}(t)$ &&-0.033&-0.101   &  - 0.169& &      -0.034&-0.101   &  - 0.165      \\
%\hline
\end{tabular}
\end{center}
}
\end{table}

\begin{table}[ht!]
\caption{Simulation results concerning the simple estimator  $\hat \delta_1(t,1)$ and the one based on the efficient influence function, $ \hat \delta_{1e}(t,1)$. 
We let $\mbox{see}_{ei}$ denote the standard error estimate based on the squared efficient influence function and $\mbox{see}_{b}$ denotes the standard error estimate based 250 bootstrap replicates.
Each entry in the table is based on 1000 replicates.
 \bigskip}

{\small
\begin{center}
\begin{tabular}{|rl|ccccc|}
\hline
&  & $t=1$ & $t=3$ & $t=5$ & $t=7$ & $t=9$  \\

\hline
A1& True $  \delta_1(t,1)$  & -0.052&-0.128& -0.178& -0.210& -0.231    \\
& Mean $ \hat \delta_1(t,1)$  & -0.052& -0.128& -0.177& -0.209& -0.229  \\
& Mean $ \hat \delta_{1e}(t,1)$   &  -0.052& -0.128& -0.178& -0.209& -0.229  \\
& sd ($  \hat\delta_1(t,1)$) &  0.014& 0.024& 0.029& 0.032& 0.034 \\
& sd ($  \hat\delta_{1e}(t,1)$)  &  0.020& 0.029& 0.033& 0.035& 0.036  \\
& $\mbox{see}_{ei}$ ($  \hat\delta_{1e}(t,1)$)  & 0.020 &0.030 &0.035& 0.037 &0.038 \\
& $\mbox{see}_{b}$ ($  \hat\delta_{1e}(t,1)$)  & 0.020& 0.029& 0.033& 0.035& 0.036 \\
A2& True $  \delta_1(t,1)$  & -0.052&-0.128& -0.178& -0.210& -0.231     \\
& Mean $ \hat \delta_1(t,1)$  & -0.052 &-0.127& -0.176 &-0.208& -0.228 \\
& Mean $ \hat \delta_{1e}(t,1)$   & -0.052& -0.127 &-0.176 &-0.207& -0.227    \\
& sd ($  \hat\delta_1(t,1)$) & 0.015& 0.028 &0.034 &0.039 &0.042   \\
& sd ($  \hat\delta_{1e}(t,1)$)  &  0.020& 0.033& 0.038 &0.042 &0.044   \\
& $\mbox{see}_{ei}$ ($  \hat\delta_{1e}(t,1)$)  &  0.020 &0.032& 0.038& 0.042& 0.046  \\
& $\mbox{see}_{b}$ ($  \hat\delta_{1e}(t,1)$)  &  0.020& 0.031& 0.038& 0.042& 0.044   \\
B1& True $  \delta_1(t,1)$  & -0.052&-0.128& -0.178& -0.210& -0.231    \\
& Mean $ \hat \delta_1(t,1)$ & -0.052& -0.127& -0.176& -0.208& -0.228 \\
& Mean $ \hat \delta_{1e}(t,1)$  & -0.052& -0.126& -0.176& -0.207& -0.228 \\
& sd ($  \hat\delta_1(t,1)$) & 0.014& 0.025& 0.031& 0.034 &0.037  \\
& sd ($  \hat\delta_{1e}(t,1)$) &0.024& 0.036& 0.041& 0.043& 0.046\\
& $\mbox{see}_{ei}$ ($  \hat\delta_{1e}(t,1)$)  &0.024& 0.037& 0.043& 0.046& 0.048 \\
& $\mbox{see}_{b}$ ($  \hat\delta_{1e}(t,1)$)  &0.023 &0.035& 0.040 &0.043 &0.044 \\
B2& True $  \delta_1(t,1)$   & -0.052&-0.128& -0.178& -0.210& -0.231    \\
& Mean $ \hat \delta_1(t,1)$ & -0.052& -0.127& -0.175& -0.206& -0.227  \\
& Mean $ \hat \delta_{1e}(t,1)$  &-0.052& -0.126& -0.175& -0.206& -0.227 \\
& sd ($  \hat\delta_1(t,1)$) & 0.016& 0.028& 0.037& 0.042& 0.045  \\
& sd ($  \hat\delta_{1e}(t,1)$) & 0.024& 0.038& 0.045& 0.049& 0.054\\
& $\mbox{see}_{ei}$ ($  \hat\delta_{1e}(t,1)$)  &  0.024 &0.039& 0.048& 0.052& 0.056 \\
& $\mbox{see}_{b}$ ($  \hat\delta_{1e}(t,1)$)  &  0.024& 0.038& 0.045 &0.050& 0.054 \\
C1& True $  \delta_1(t,1)$   & 0.065 &0.120 &0.120 &0.11 &0.092    \\
& Mean $ \hat \delta_1(t,1)$  &0.045 &0.093& 0.113& 0.121& 0.124 \\
%& sd ($  \hat\delta_1(t,1)$) & & & & & & & \\
& Mean $ \hat \delta_{1e}(t,1)$   &  0.064& 0.114& 0.118& 0.105& 0.091  \\
%& sd ($  \hat\delta_1(t,1)$) & & & & & & &  \\
& sd ($  \hat\delta_{1e}(t,1)$) &0.030 &0.040 &0.042 &0.041 &0.041\\
& $\mbox{see}_{ei}$ ($  \hat\delta_{1e}(t,1)$)  & 0.030& 0.041& 0.043& 0.044& 0.043  \\
& $\mbox{see}_{b}$ ($  \hat\delta_{1e}(t,1)$)  & 0.030& 0.040 &0.042 &0.042& 0.042  \\
C2& True $  \delta_1(t,1)$  & 0.065 &0.120 &0.120 &0.110 &0.092     \\
& Mean $ \hat \delta_1(t,1)$  & 0.054& 0.111 &0.134& 0.144& 0.147   \\
%& sd ($  \hat\delta_1(t,1)$) & & & & & & & \\
& Mean $ \hat \delta_{1e}(t,1)$   & 0.066& 0.117& 0.122& 0.111& 0.099  \\
%& sd ($  \hat\delta_1(t,1)$) & & & & & & &  \\
& sd ($  \hat\delta_{1e}(t,1)$) & 0.032& 0.043& 0.047 &0.049 &0.050\\
& $\mbox{see}_{ei}$ ($  \hat\delta_{1e}(t,1)$)  &  0.031& 0.044& 0.048& 0.051& 0.052   \\
& $\mbox{see}_{b}$ ($  \hat\delta_{1e}(t,1)$)  &  0.030& 0.043& 0.048& 0.050 &0.051  \\
\hline
%\hline
\end{tabular}
\end{center}
}
\end{table}

  \begin{center}
\begin{figure}[ht!]
  \includegraphics[width=16cm,height=11cm]{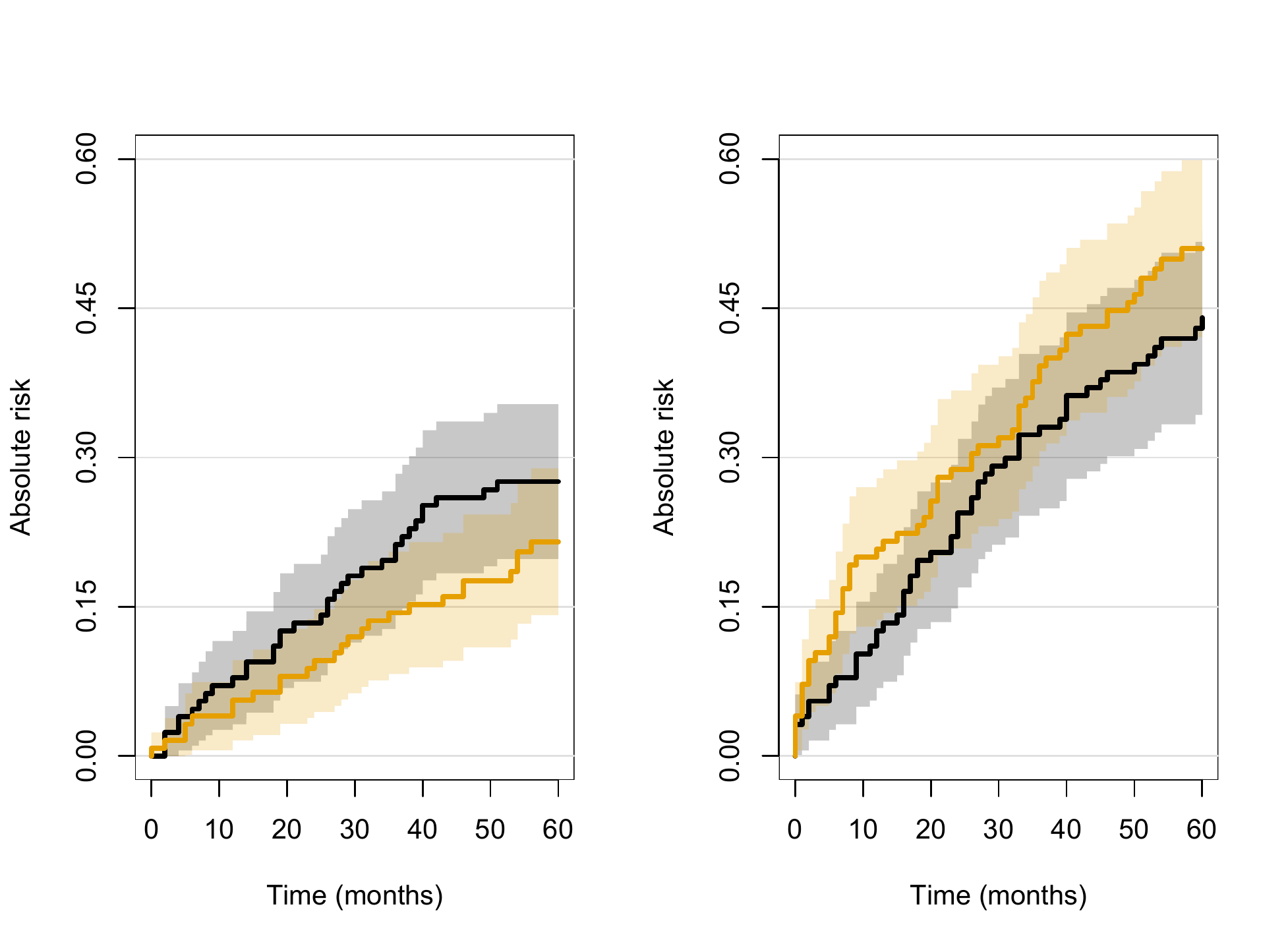}
  \caption{Results from the prostate cancer data example. The curves describe the cumulative incidence of prostate cancer death (left panel) and death due to other causes (right panel). The black curves denote the placebo arm while the orange curves denote the treatment arm, and the shaded areas are 95\% confidence bands.}
 \end{figure}
  \end{center}

  \begin{center}
\begin{figure}[ht!]
  \includegraphics[width=16cm,height=10cm]{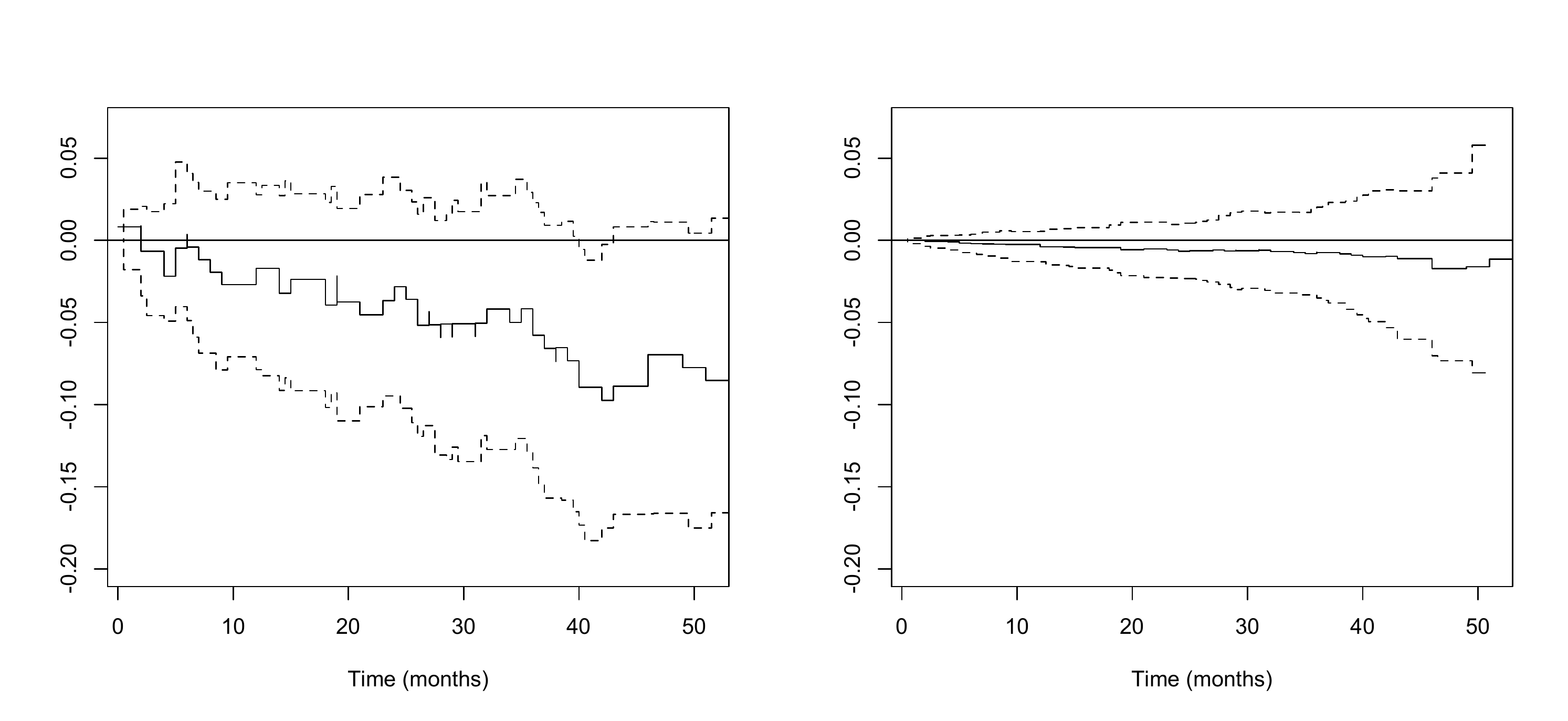}
  \caption{Separable effects in the prostate cancer data application. Left panel shows $\hat\delta_{1e}(t,0)$ (full curve) against time (months) with dashed lines showing 95\% confidence bands using the non-parametric bootstrap. Right panel shows the corresponding separable indirect effect (full curve)  with dashed lines showing 95\% confidence bands using the non-parametric bootstrap.
  }
 \end{figure}
  \end{center}
  
\noindent

\end{document}